\documentclass[a4paper,11pt]{article}
\pdfoutput=1 

\usepackage{jinstpub} 

\usepackage{caption}
\usepackage{subcaption}

\title{\boldmath High Rate Studies of the ATLAS sTGC Detector and Optimization of the Filter Circuit on the Input of the Front-End Amplifier}

\author[a,1]{Siyuan Sun\note{Corresponding author}}
\author[b,2]{Luca Moleri \note{Corresponding author, currently at the Weizmann Institute of Science, Rehovot, Israel}}
\author[c,3]{Gerardo Vasquez\note{Corresponding author}}
\author[d]{Peter Teterin}
\author[a]{Sabrina Corsetti}
\author[a]{Liang Guan}
\author[e]{Benoit Lefebvre}
\author[b]{Enrique Kajomovitz}
\author[f]{Lorne Levinson}
\author[b]{Nachman Lupu}
\author[c]{Rob McPherson}
\author[b]{Alexander Vdovin}
\author[a]{Rongkun Wang}
\author[a]{Bing Zhou}
\author[a]{Junjie Zhu}

\affiliation[a]{University of Michigan, Ann Arbor, MI, USA}
\affiliation[b]{Technion, Haifa, Israel}
\affiliation[c]{University of Victoria, Victoria, BC, Canada}
\affiliation[d]{National Research Nuclear University MEPhI, Moscow, Russia}
\affiliation[e]{TRIUMF, Vancouver, BC, Canada}
\affiliation[f]{Weizmann Institute of Science, Rehovot, Israel}

\emailAdd{sssun@umich.edu}
\emailAdd{g.vasquez@cern.ch}
\emailAdd{luca.moleri@weizmann.ac.il}

\abstract{The Large Hadron Collider (LHC) at CERN is expected to be upgraded to the High-Luminosity LHC (HL-LHC) by 2029 and achieve instantaneous luminosity around $5-7.5 \times 10^{34}$ cm$^{-2}$ s$^{-1}$.  This represents a more than 3-4 fold increase in the instantaneous luminosity compared to what has been achieved in Run 2. The New Small Wheel (NSW) upgrade is designed to be able to operate efficiently in this high background rate environment. In this article, we summarize multiple performance studies of the small-strip Thin Gap Chamber (sTGC) at high rate using nearly final front-end electronics.  We demonstrate that the efficiency versus rate distribution can be well described by an exponential decay with electronics dead-time being the primary cause of loss of efficiency at high rate. We then demonstrate several methods that can decrease the electronics dead-time and therefore minimize efficiency loss.  One such method is to install either a pi-network input filter or pull-up resistor to minimize the charge input into the amplifier.  We optimized the pi-network capacitance and pull-up resistor resistance using the results from our measurements. The results shown here were not only critical to finalizing the components on the front-end board, but also are critical for setting the optimal operating parameters of the sTGC detector and electronics in the ATLAS cavern.}


\begin{document}
\maketitle
\flushbottom

\section{Introduction}
\label{sec: intro}

The Large Hadron Collider (LHC) at CERN is expected to be upgraded to the High-Luminosity LHC (HL-LHC) by 2029.  As it's name suggests, the HL-LHC is expected to achieve instantaneous luminosities around $5-7.5 \times 10^{34}$ cm$^{-2}$ s$^{-1}$.  This represents a more than $3-4$ fold increase in the instantaneous luminosity compared to what has been achieved in Run 2.  

An increase in the instantaneous luminosity corresponds to a similar increase in background rates~\cite{HitRate}; mostly from low energy photons and neutrons. The latter, having a large elastic cross section, bounce in the detector cavern and generate random background signal long (order of microseconds) after their production in the collision. We refer to this collective diffuse gas of long-lived neutral particles as "cavern background". Previous measurements of cavern background using the ATLAS monitored drift tube (MDT) system have found that up to 65\% of hits in the MDTs come from a long-lived component with a half-life of 50 microseconds~\cite{NSW_TDR}.

At HL-LHC luminosities, we expect background rates of up to 15 kHz/cm$^2$ in the end-cap region of the ATLAS muon spectrometer~\cite{NSW_TDR}. This region of the muon spectrometer, called the Small Wheel, is the region that is closest to the proton-proton interaction point and closest to the beam-line.

\begin{figure}[h!]
\centering 
\includegraphics[width=.6\textwidth,trim=0 6 0 0,clip]{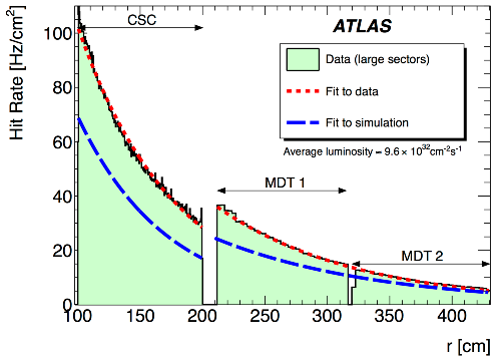}
\caption{\label{fig:LHCrate} The measured background rate vs $r$ in the previous small wheel detectors. The background hit rate increases exponentially with decreasing $r$. HL-LHC instantaneous luminosity are predicted to be up to $5-7.5 \times 10^{34}$ cm$^{-2}$ s$^{-1}$ compared to the $9.6 \times 10^{32}$ cm$^{-2}$ s$^{-1}$ luminosity in this plot. Plot taken from~\cite{NSW_TDR}. }
\end{figure}

As such, we expect the highest rate of background, with the background rate falling exponentially with $r$, the distance to the beam line, decreases as shown in Fig.~\ref{fig:LHCrate}~\cite{NSW_TDR}.
The background rate, scaled with detector area, corresponds to a maximum of 1 MHz per detector channel.  Current detectors cannot operate efficiently at such high rates and as such, necessitate an upgrade~\cite{NSW_TDR}.  

The New Small Wheel (NSW) upgrade is designed to be able to operate efficiently in this high background rate environment. The small-strip Thin Gap Chamber (sTGC) is one of two detector technologies used. Both the detector and the electronics need to operate efficiently in this high background environment. 

In this article, we summarize the results of multiple tests of sTGC detector and electronic performance at high rates.  
Section \ref{sec: TB October} covers the results of the first test beam of the sTGC detector in a high background rate environment in October 2018.  This test beam is the first measurement of detector efficiency vs background rate and found that the drop in efficiency at high rates can be fully parameterized by a exponential decay with a single parameter: the electronics dead-time. 

Section \ref{sec: GIF++ analysis} covers the result of a second test beam in a high background environment in January - March 2019.  This test beam focused on minimizing the electronics dead-time and the efficiency loss at high rate.  Different input charge filters that limit the charge input into the amplifier were tested.  We also test different front-end ASIC configurations that incur less dead-time.  



Measurements of the relationship between efficiency at high rates, dead-time, and input charge per hit made possible to set the optimal value for the charge filter that is placed in front of every sTGC pad input channel. The filter decreases the input charge per hit and minimizes the incurred dead-time while maintaining muon signal strength.

The present measurements also provide a more complete understanding of the expected sTGC efficiency at HL-LHC rates and recommendation for sTGC electronics settings when operating at high rates.

\subsection{The sTGC Detector}
\label{sec: intro_sTGC}

\begin{figure}[h!]
\centering
\includegraphics[width=0.7\textwidth]{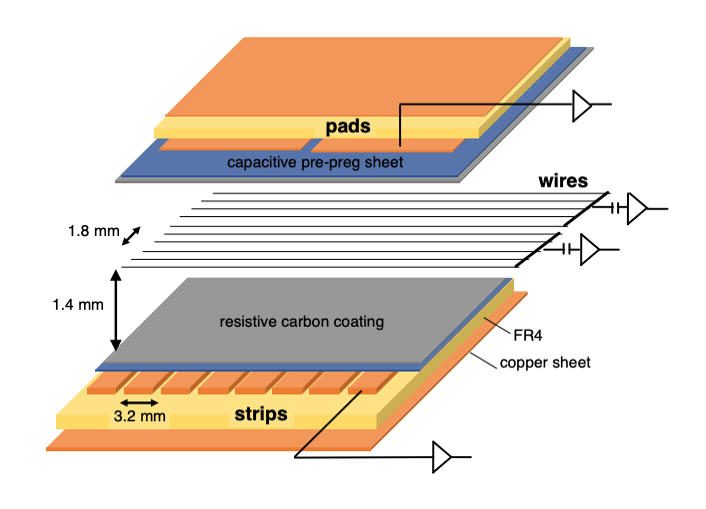}
\caption{\protect A diagram of the sTGC detector.  The wire anode is sandwiched between the pad and strip cathode planes. Taken from~\cite{sTGC_Proceeding}. }
\label{fig: sTGC_Detector}
\end{figure}

The small-strip Thin Gap Chamber (sTGC) is a type of gaseous detector with a geometry configuration of a multi-wire proportional chamber  but working in a high gain mode (limited proportional region). The chamber is composed of a plane of gold plated tungsten wire anodes sandwiched by two cathode planes as shown in Fig.\ref{fig: sTGC_Detector}. 

The gas gaps between the cathode planes are filled with a CO$_2$+n-pentane gas mixture in a ratio of 55:45, respectively. High voltage (typically 2800~V) is applied between the cathode and anode forming a strong electric field. Energetic charged particles ionize the gas molecules as they pass through the detector. The liberated primary electrons and ions will drift along the electric field lines to the anode and cathodes, respectively.  The electric field is most concentrated around the anode wires.  The primary electrons will be multiplied in an electron avalanche as they approach the anode wires due to the strong electric field around them.  The motion of the electron-ion pairs formed in the avalanche induces a signal in the cathode and anode according to the Shockley–Ramo theorem \cite{Shockley, Ramo}.

The cathodes are composed by two planes. A graphite-epoxy layer coated on FR4 faces the wires and serves for slow charge evacuation. The resistive layer and cathode-anode gap form an RC circuit that limits the speed of charge dispersion to ground, reducing the intensity of occasional discharges. The resistivity of the graphite is in the order of 100 k$\Omega$ to M$\Omega$. On the FR4 external side a copper plane is segmented either in pads or strips, which serve as signal pickup electrodes. 

Important sTGC detector parameters are given in table \ref{tab: sTGC_Detector}.

\begin{table}
\centering
    \begin{tabular}{p{6cm}|p{2cm}}
    \hline \hline
    \multicolumn{2}{c}{sTGC Geometrical and physical parameters}\\
    \hline
    wire-to-cathode pitch& 1.4 [mm]\\
    \hline
    wire-to-wire pitch & 1.8 [mm]\\
    \hline
    wire diameter & 50 [$\mu$m]\\
    \hline
    strip pitch & 3.2 [mm]\\
    \hline
    Graphite layer thickness & 10 [$\mu$m]\\
    \hline
    Q1 Graphite surface resistivity & 100 k$\Omega/\square$\\    \hline
    Q2/Q3 Graphite surface resistivity & 200 k$\Omega/\square$\\
    \hline \hline
\end{tabular}
    \caption{Detector parameters}
    \label{tab: sTGC_Detector}
\end{table}

\begin{figure}[h!]
\centering
\includegraphics[width=0.9\textwidth]{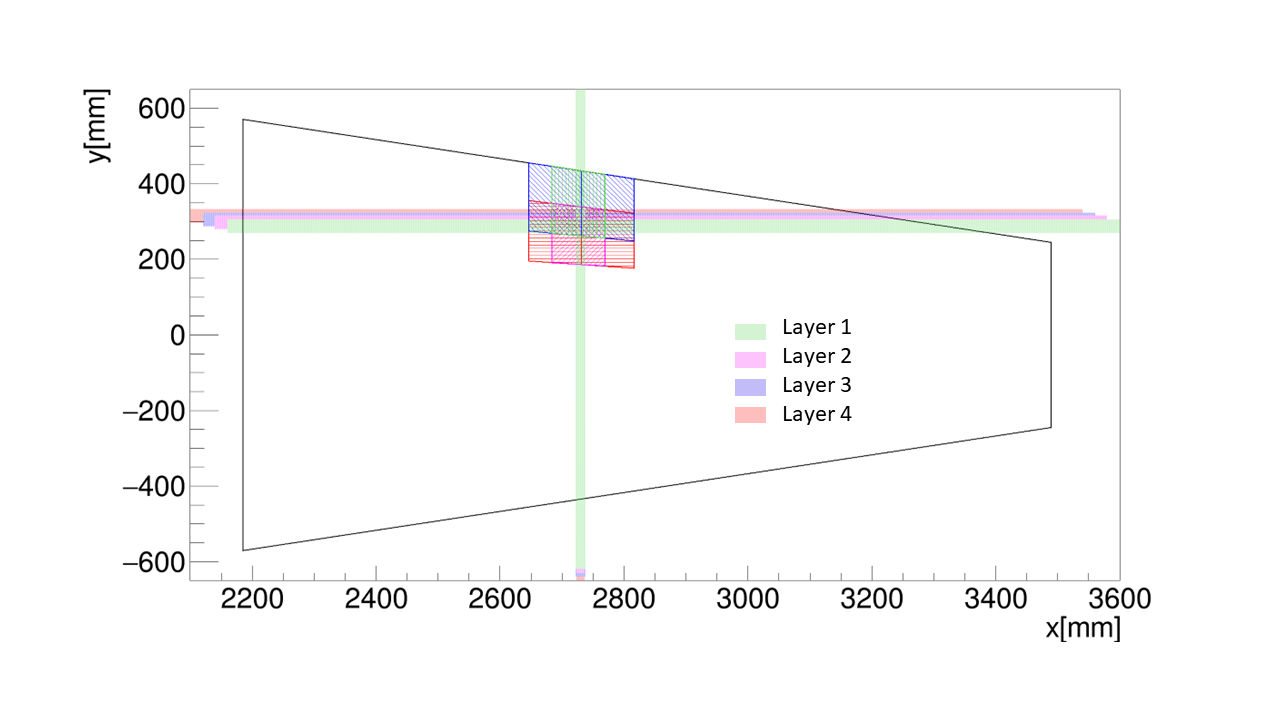}
\caption{\protect The hits generated in all 4 layers of the detector from a muon. Each color represent hits in the pads, wires, and strips in each layer of the quadruplet.  The overlapping pads, strips and wires allow the identification of the location of the muon in the 2D plane. }
\label{fig: sTGC_event_display}
\end{figure}

Four sTGC gaps are stacked together to form a quadruplet. This allows implementing different logic coincidence schemes to optimize efficiency and background rejection.  An event display of the typical signal a muon generates in the quadruplet can be seen in figure \ref{fig: sTGC_event_display}.  The overlapping pad hits are used to form a quick region of interest (ROI) for the online trigger.  The segment is then reconstructed in the ROI from the clusters of hits on the strips.  Precise location information is derived from the centroid of the strip clusters.  The segment information is than forwarded to the back-end trigger electronics to be integrated with trigger information from other muon detectors.  Hit information from pads, wires, and strips are also used in offline muon track reconstruction.

\begin{figure}[h!]
\centering
 \begin{subfigure}[b]{0.45\textwidth}
\includegraphics[width=\textwidth]{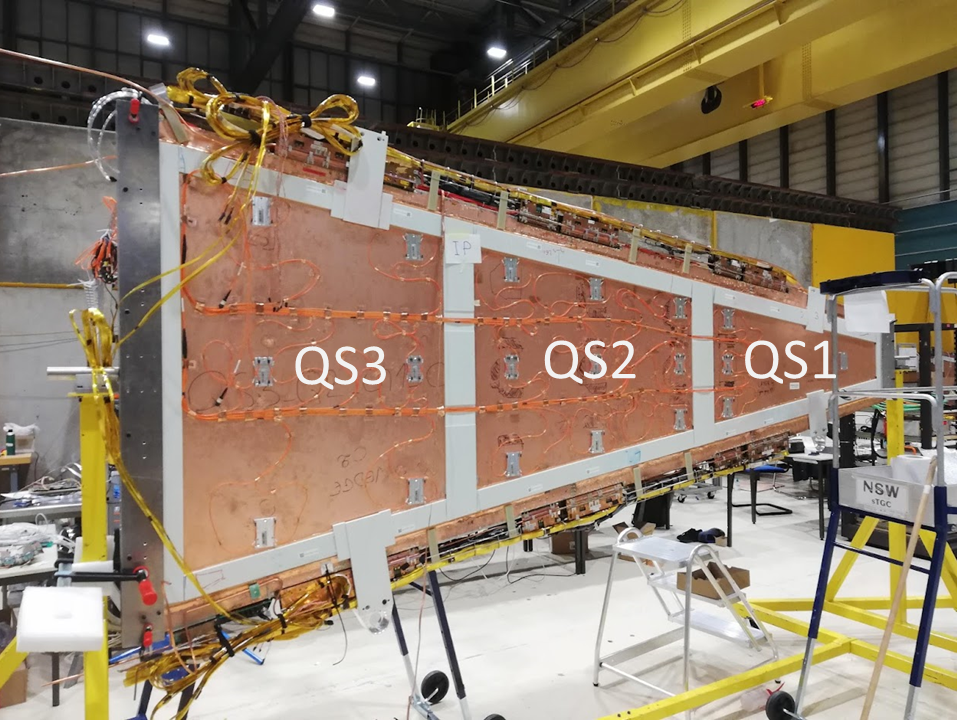}
\caption{}
\label{subfig: SmallWedge}
 \end{subfigure}
\qquad
 \begin{subfigure}[b]{0.45\textwidth}
\includegraphics[width=\textwidth]{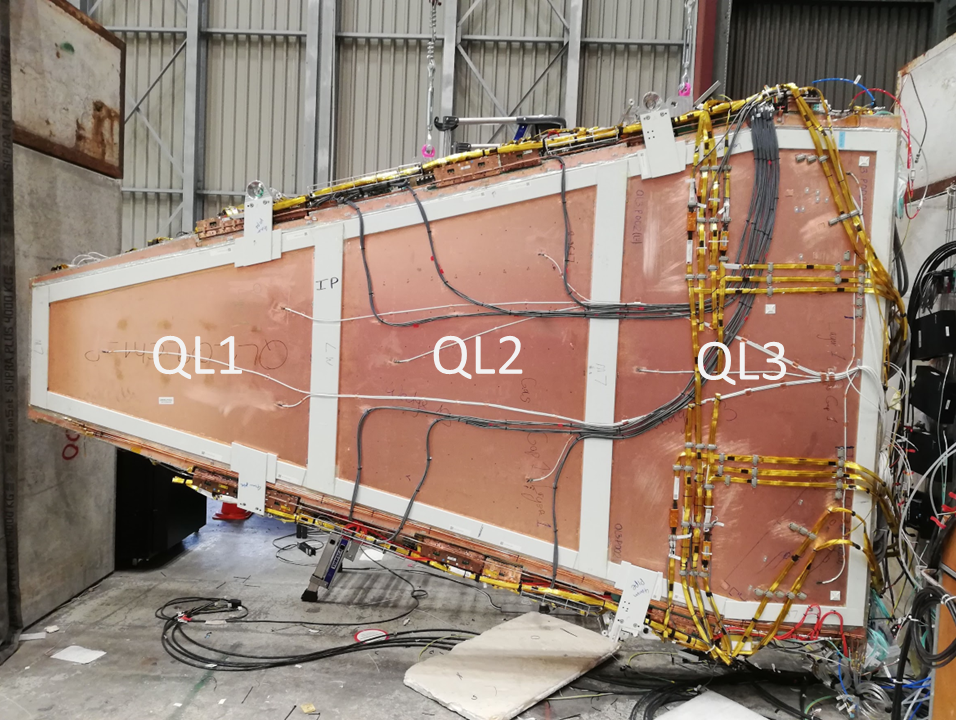}
\caption{}
\label{subfig: LargeWedge}
\end{subfigure}
\caption{ \ref{subfig: SmallWedge} A fully assembled sTGC small wedge \ref{subfig: LargeWedge} A fully assembled sTGC large wedge. The Q1, Q2 and Q3 in each wedge differ greatly in pad sizes because the expected background rate decreases exponentially with increasing distance from the beam line. }
\label{fig: wedges}
\end{figure}

Three quadruplets are mechanically assembled to form wedges.  Two different wedge type exists, large and small, as shown in figure \ref{fig: wedges}.  The smallest quadruplets, QS1 and QL1, are located in the inner most radius.  The largest quadruplets, QS3 and QL3, are located in the outer most radius.  The size of the sTGC pads differ greatly between each quadruplet type because of the difference in expected background rate per unit area with radius R.  QS1 and QL1 pads are typically around 40-160 cm$^2$.  QS2 and QL2 pads range from 150 to 400 cm$^2$.  QS3 and QL3 pads range from 350 to 550 cm$^2$. In this work we describe results from testing two individual quadruplets models, namely QL1 and QS2. Testing QS1 chambers would
also be very useful, but those chambers were not available for investigation at the time of the present work.

\subsection{The VMM3a Front-End ASIC}
\label{sec: intro_VMM}

\begin{figure}[h!]
\centering
\includegraphics[width=0.7\textwidth]{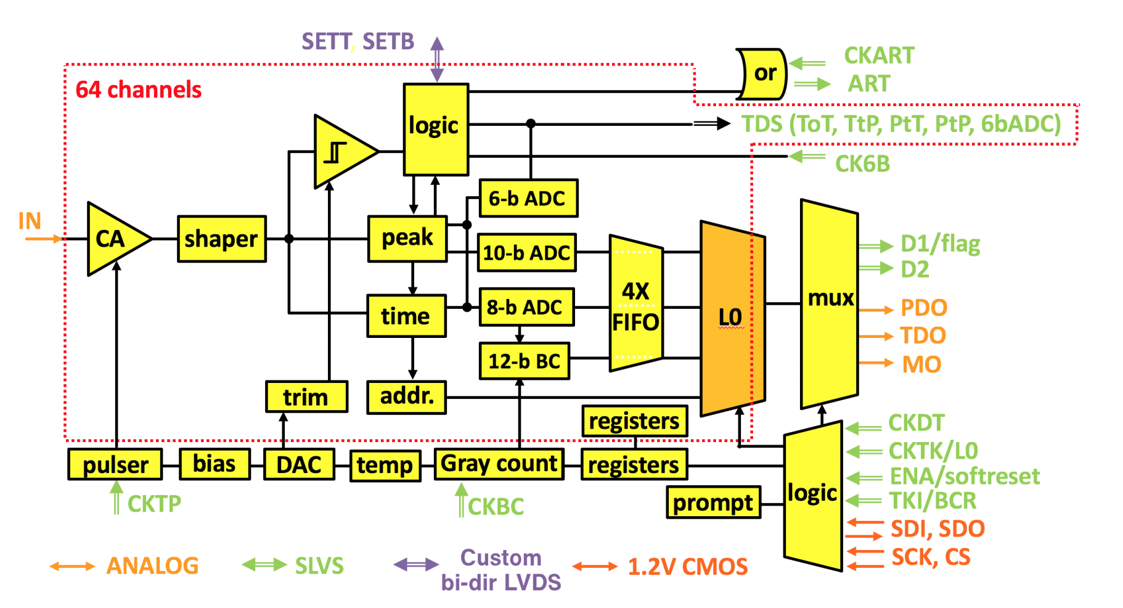}
\caption{\protect Block diagram of the VMM3a ASIC, the primary front-end ASIC for the New Small Wheel. Taken from~\cite{VMM_Spec} }
\label{fig: VMM_BlockDiagram}
\end{figure}

The VMM3a is a mixed-signal front-end ASIC used for the New Small Wheel~\cite{VMM_Spec}. A block diagram of the VMM3a can be seen in Fig.~\ref{fig: VMM_BlockDiagram}. It features 64 channels, each with an amplifier, shaper, discriminator and a peak detector. For each VMM channel, precision charge and arrival time information of the detected signal is digitized with a 10-bit Analog-Digital Converter (ADC) and a Time-Amplitude Converter (TAC) followed by an 8-bit ADC, respectively. The digitized channel hit data are buffered inside the chip and could be read out upon the reception of external trigger signals. The digital logic on the ASIC performs the matching of the trigger signal with all channel hits which fall within an adjustable timing window, of up to 200 ns. In the application within the ATLAS experiment, hit data associated with a trigger are built into an event data packet and sent via a high-speed serial data port to the Readout Controller ASIC (ROC)\cite{ROC_Spec}, which aggregates data from multiple VMMs. In this readout chain, a VMM channel operates continuously and is ready to process a new hit once the previous charge signal falls below the threshold and the channel digitization resets. 

In parallel to the readout chain mentioned above, the VMM ASIC is also equipped with direct output pins from which a less precise but prompt channel hit information could be sent out for fast trigger purposes. The direct output for each VMM channel can be configured to send out either a timing pulse or a (lower-resolution) 6-bit ADC charge data. The direct output information could be utilized to take a fast online trigger decision and could be subsequently used to select precision hit information for offline track reconstructions of charged particles. The direct outputs in the trigger chain send less information about each hit than the readout chain, according to different operational modes. For example, time-over-threshold (ToT) pulses could be sufficient for identifying coincidences among multiple detector layers. This output mode is favoured for sTGC pad channels in the ATLAS first-level hardware-based triggering. Similarly, the 6-bit ADC data output is favourable for triggering with sTGC strip channels, where prompt charge information could be used to determine the signal centroid. This would allow to perform precision particle tracking at the trigger level. 


The dead-time generated after a hit for a VMM3a channel is of special importance to the present measurements and, as will be demonstrated later, of prime importance to the efficiency loss at high background rate. As such, careful attention must be paid to the recovery of the VMM3a amplifier and the channel reset procedure of its various ADCs after detecting a hit.  

\begin{figure}[h!]
\centering
\includegraphics[width=0.7\textwidth]{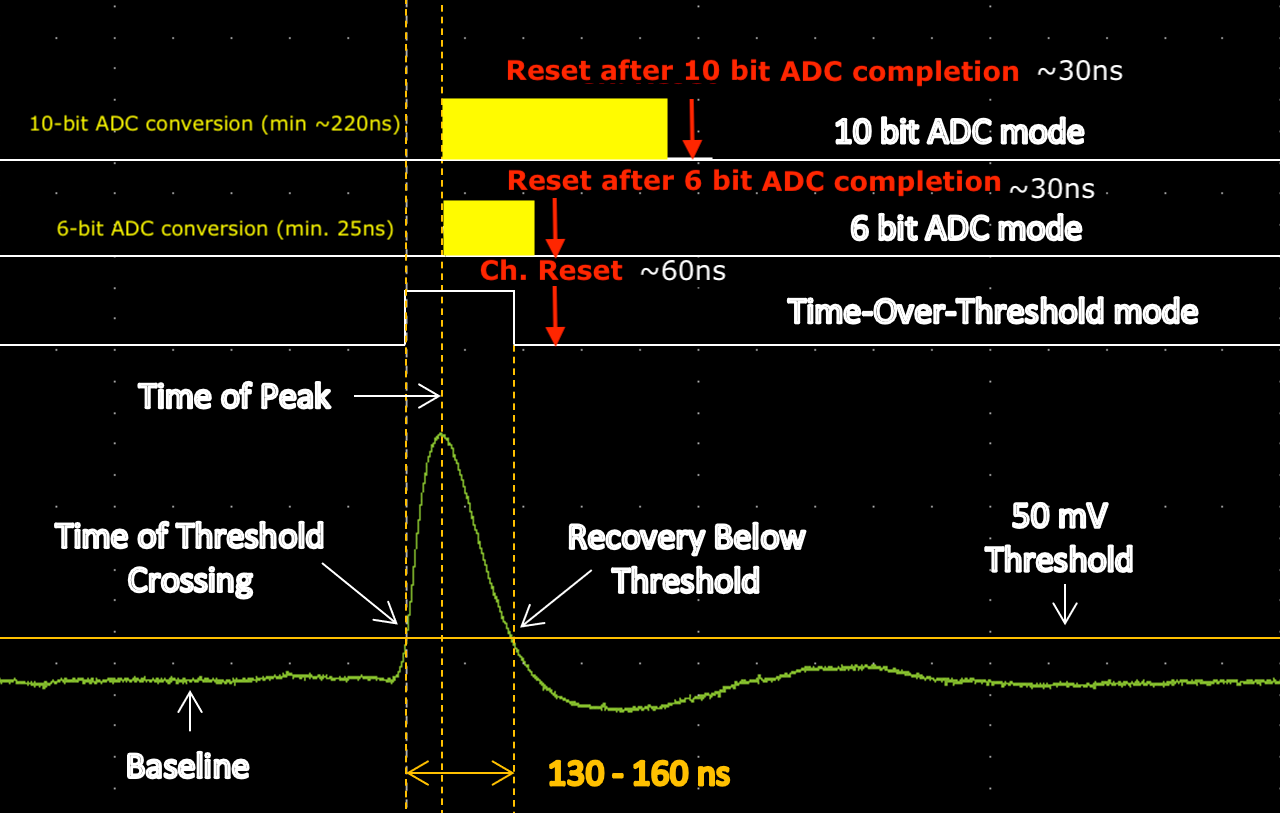}
\caption{VMM3a analog response to a typical sTGC signal produced by a high energy muon and the channel digitization and reset procedure in different operation modes. The 10-bit ADC mode is  typically limited by the 220 ns digitization time of the 10-bit ADC. The time-over-threshold (ToT) and 6-bit ADC mode are both usually limited by the recovery of the analog pulse below threshold after a hit.} 
\label{fig: VMM_Reset}
\end{figure}

Figure \ref{fig: VMM_Reset} illustrates the VMM3a channel reset procedure in different operational modes. The peak of the VMM3a analog output is reached after a time, which depends on the ASIC peaking time and on the detector signal shape. Studies using a detector emulation method as input to the ASIC show that the peaking time for sTGC signals is around $35-40$ ns and $60-70$ ns for a signal with VMM3a peaking time setting of 25 ns and 50 ns, respectively~\cite{DetEmulation}. Both 10-bit ADC and 6-bit ADC start digitization a few nanosecond following the signal peak detection.
The digitization process of the 10-bit ADC takes about $220-400$ ns to complete. When the VMM3a is configured to reset a channel upon completion of 10-bit ADC conversion, the channel will take another 30 ns at the end of 10-bit ADC conversion to reset the channel. In this mode, a minimum reset time of 250 ns~\cite{VMM_Spec} is expected after the peak detection. This VMM operational mode was kept as the default in the presented study.  

In a different configuration, the VMM3a channel could be forced to reset as soon as the fast 6-bit ADC digitization is complete. The 6-bit ADC conversion time is configurable to be between $25-68.75$ ns. The same 30 ns is needed after the completion of the 6-bit ADC digitization to reset the channel. This gives a theoretical minimum dead time after threshold-crossing of $\sim$120 ns, for a typical VMM3a peaking time configuration of 50 ns. However, this is an ideal scenario which underestimates the dead-time, since the analog pulse must also return to baseline before the VMM channel is ready for another hit. For a moderate amount of charge, such as 6 pC, the decay time for the analog pulse can be several times the selected peaking time due to the semi-Gaussian shaper design. This combined with a signal rise time of $60-70$ ns gives a single hit processing time no less than $\sim$140 ns. For large charge deposits, due to the tail of the energy deposited by a minimum ionizing particle or to neutral particles such as photons and long-lived neutrons, the recovery time can be as long as 1-2 $\mu$. In this case, the time which the analog pulse takes to return below threshold is therefore the limiting factor regardless of the ADC digitization time.

Finally, the VMM ASIC could be configured to force a channel reset at the end of the time-over-threshold output. In this case, the channel reset occurs about 60 ns after the trailing edge of the ToT pulse for moderate input charge. The channel will not be waiting for the completion of the ADC and therefore the charge information will be lost. Also in this case, the limiting factor is the recovery of the charge amplifier response to analog pulses. In a typical sTGC application with a VMM3a peaking time of 50 ns and a reasonable threshold of 30-50 mV, the total channel reset time is expected to be no less than ~200 ns, taking into consideration the semi-Gaussian pulse-shape output from the shaping amplifier.


As we shall see, the three modes of channel reset offer different benefits and drawbacks. The channel reset upon 10-bit ADC completion provides precise charge information that is needed by sTGC readout strips to accurately reconstruct hit positions. However, this introduces a long dead-time limiting the detector system efficiency and performance at high rate. Forcing a reset of VMM3a channels upon completion of prompt output (6-bit ADC or ToT pulses) could reduce the overall dead-time. However, this would sacrifice the precision 10-bit charge and 8-bit fine timing information in the readout chain. Therefore, it is important to note the interplay among different ADC or timing output reset modes and its impact on the readout and trigger chain operations. For the application of sTGC in the ATLAS experiment, it is natural to choose a channel reset upon ToT completion for pads to minimize the dead-time, as they have much larger areas than strips and have to tolerate high per-channel counting rate. For sTGC strips, the VMM3a has to wait for the completion of 10-bit ADC in the readout chain while simultaneously sending 6-bit ADC for triggering purposes.

\subsection{The $\pi$-Network Filter and Pull Up Resistor on the VMM input}
\label{sec: intro_PiNetwork}

\begin{figure}[h!]
\centering
\includegraphics[width=0.7\textwidth]{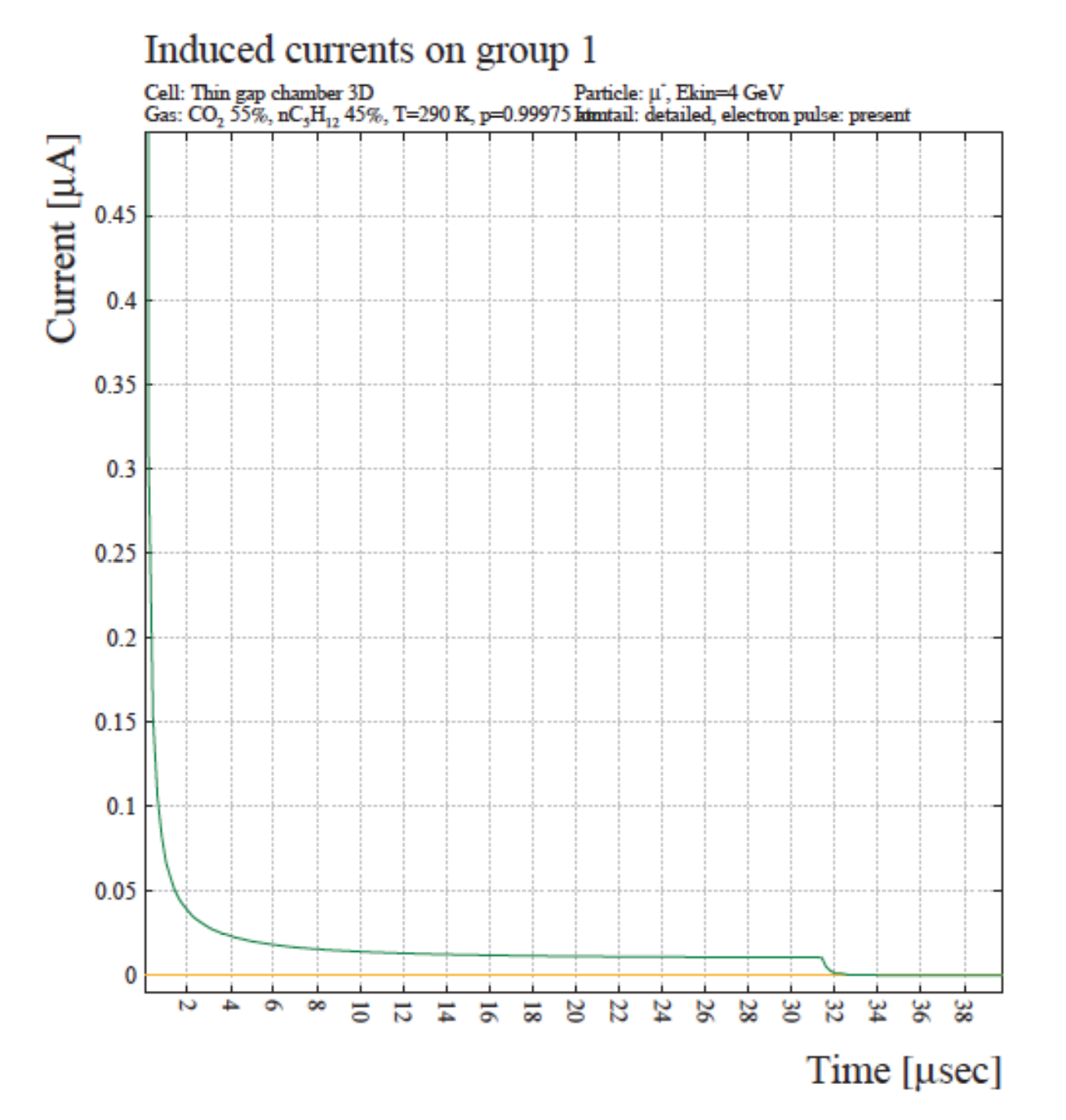}
\caption{Simulated distribution of the signal for the sTGC detector.  Y-axis is current on the detector channel and y-axis is time in microseconds. The fast immediate electron component and the slow ion component of the signal can be seen. The very slow component of negative polarity due to the charge flow across the resistive layer is seen here as it happens on a much larger (msec) time scale. Taken from~\cite{NSW_TDR}. }
\label{fig: sTGC_Signal}
\end{figure}

\begin{figure}[h!]
\centering
\includegraphics[width=0.7\textwidth]{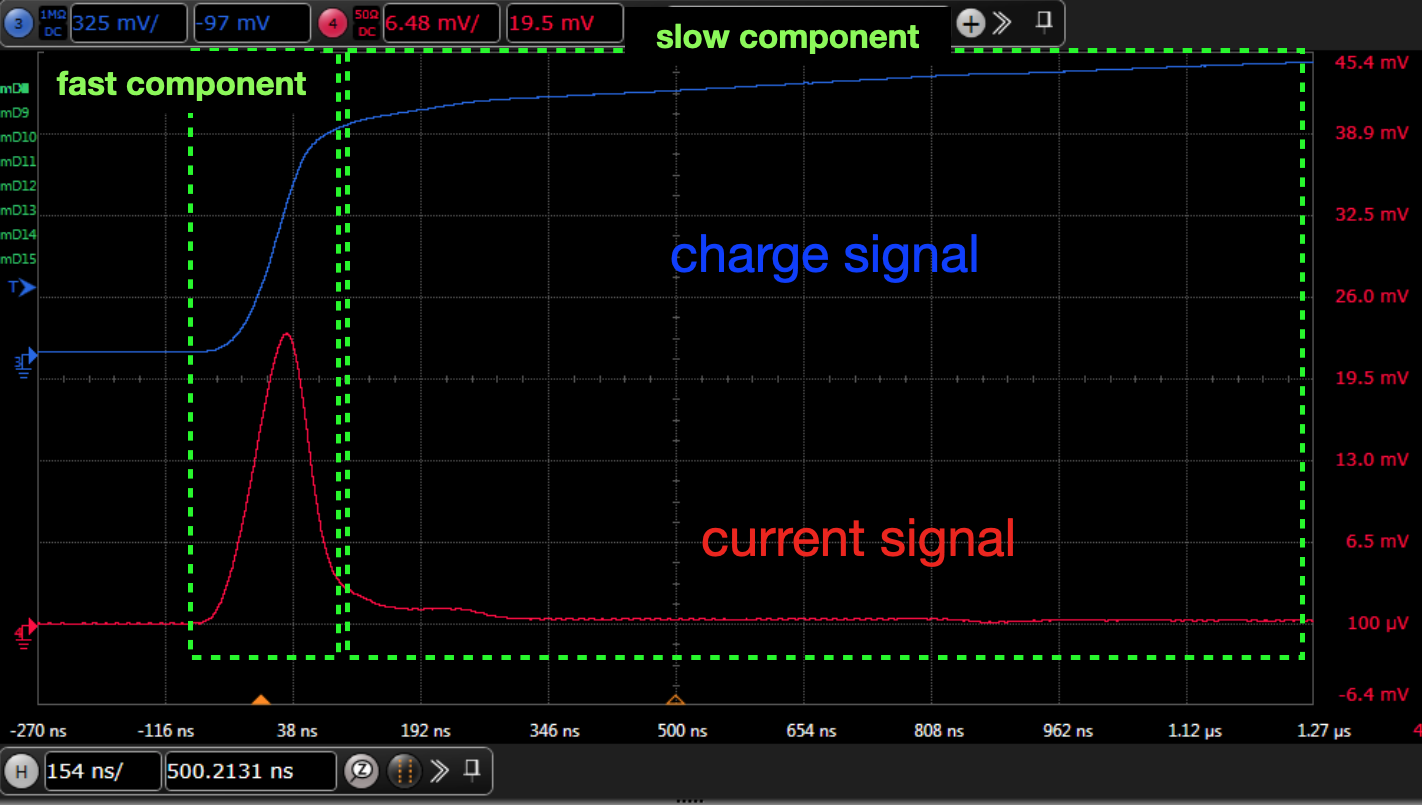}
\caption{The average sTGC integrated charge signal shape (in blue) and its derivative (in red, corresponding to the current signal). Taken from~\cite{DetEmulation}
}
\label{fig: emulator signal}
\end{figure}

The signal from the sTGC detector consists of three components: A fast ($\sim$20 ns) signal of a positive polarity due to the drifting of avalanche electron towards the wire. A slow (order 10 $\mu$s) signal of positive polarity due to the drifting of the avalanche-ions towards the cathodes, and a very slow (msec) component of negative polarity due to the charge flow across the resistive layer. A plot of the simulated input current from an sTGC channel is given in figure \ref{fig: sTGC_Signal} (the third component is not visible here)~\cite{VMM_Spec}. A measurement of average sTGC current and charge signals is shown in figure \ref{fig: emulator signal} \cite{DetEmulation}, showing the fast and slow components.


At high rates, the long ion tail distribution from many hits will be superimposed upon one another to form a near constant input current into the amplifier.  This is especially true for the sTGC detector which expect to have a high background rate of mainly highly ionizing long-lived neutral particles such as low energy neutrons.  The VMM has a baseline recovery circuit to counter-act this low frequency input and maintain the stability of the baseline at high rates. 

\begin{figure}[h!]
\centering
 \begin{subfigure}[b]{0.45\textwidth}
\includegraphics[width=\textwidth]{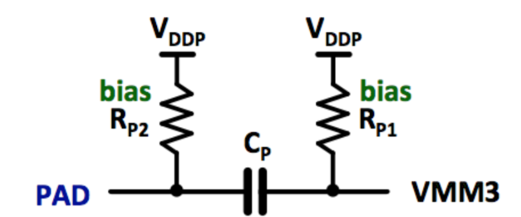}
\caption{}
\label{subfig: PiNetwork}
 \end{subfigure}
\qquad
 \begin{subfigure}[b]{0.45\textwidth}
\includegraphics[width=\textwidth]{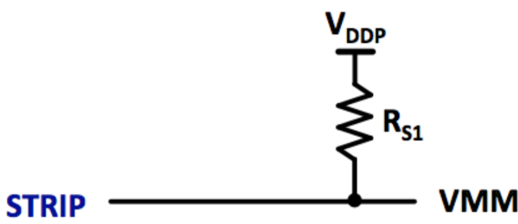}
\caption{}
\label{subfig: StripPullUp}
\end{subfigure}
\caption{ \ref{subfig: PiNetwork} The $\pi$-Network filter placed in between a sTGC pad and the input to the VMM amplifier.  \ref{subfig: StripPullUp} The pull-up resistor placed in between a sTGC strip and the input to the VMM amplifier.  }
\label{fig: PiNetworkDiagram}
\end{figure}

A single cavern background hit can deposit as much as 50 pC of charge in one sTGC pad according to simulation \cite{VMM_Spec}, which will momentarily saturate the ASIC's amplifier. For this reason, an additional input filter circuit called the $\pi$-network was designed to attenuate the signal of the sTGC pad channels. The pad $\pi$-network circuit diagram can be seen in figure \ref{fig: PiNetworkDiagram}. The $\pi$-network acts as a charge divider, splitting the charge left in the detector channel with that of it's own capacitor. At the same time, the two bias resistors connected to the 1.2V power rail act as a source of current to counter-act the constant input current due to the ion tail, thereby acting as an aid to the VMM's baseline recovery circuit.  

 The $\pi$-network and detector channel capacitors are in series and the charge accumulated will be divided among the two according to their capacitance.  In addition to reducing detector noise and background hits, the $\pi$-network capacitor also attenuates the muon signal. The capacitance and attenuation must be optimized to maintain the size of the muon signal but limit the amount of charge from highly ionizing photons and neutrons that can saturate the VMM's amplifier.  
 
 An estimate of the attenuation factor of the $\pi$-network is approximated by equation \ref{eqn: pi-attenuation}, where $C_{p}$ and $C_{detector}$ are the capacitances of the $\pi$-network and of the detector, respectively.  The default value of $C_p$ was 100pF.

\begin{equation}
    f = \frac{C_\textup{p}}{C_\textup{p}+C_\textup{detector}}
    \label{eqn: pi-attenuation}
\end{equation}

We expect the charge from a single hit to be split across more than 3 strips and so the input charge per strip channel is less than that of a pad channel~\cite{NSW_TDR}.  Simulation of the ASIC performance suggest that a charge divider such as that used for the pads is not needed here. Instead, a pull-up resistor is added to the sTGC strips channels.  The pull-up resistor is able to draw a constant counter-acting current that flows counter to that of the long-ion tail and thereby partially cancelling it. The strip pull up resistor circuit diagram can be seen in figure \ref{fig: PiNetworkDiagram}.  

A resistance that is too small will correspond to too large an input current at low rates with no long-ion tail current to cancel. The current through the resistor will itself overwhelm the baseline recovery circuit at low background rates rendering the amplifier non-functional. Therefore, an optimization of the resistance value is needed. 

The measurements in this paper explain how to quantify the effect of these circuits and how the optimal values of the $\pi$ network capacitor and the strip pull up resistor were determined.

\subsection{The Front-End Boards}
\label{sec: intro_FEB}

The sTGC uses two front-end boards called the pad front-end board (pFEB) and the strip front-end board (sFEB) to readout the sTGC pads and strips, respectively. \cite{sTGCFEB} 

The VMM3a version is used in theses tests and it is described in section \ref{sec: intro_VMM} is the primary front-end ASIC for the sTGC system each with 64 electronics channels.  A pFEB has 3 VMMs, two reading the cathode pad channels signal and 1 for the wire channels giving the option for a secondary phi-coordinate.  A sFEB has 8 VMMs for the more than 400 strip channels on each detector layer. 

Each board also has a Readout Controller ASIC (ROC)\cite{ROC_Spec} that serves to aggregate all the data packets corresponding to the same L1A trigger signal from all the VMMs on a front end board into a single event and sends the data down the line to the DAQ system.  The ROC also distributes the clock and all synchronous signals such as timing, trigger, and control (TTC) to the rest of the ASICs on the front-end board.

The Slow Control Adapter (SCA)\cite{SCA_Manual} performs all slow control tasks such as configuring all the other front-end ASICs using it's multiple SPI and I2C lines.  The SCA also features multiple ADCs that are used to monitor and calibrate the other ASICs.  

Finally, the Trigger Data Serializer (TDS)\cite{TDS_Spec} receives the direct output trigger information from VMM.  Two different TDS exists, the pad TDS receives time-over-threshold information from the VMMs on the pFEB and the strip TDS receives 6 bit ADC information from the VMMs on the sFEB.  Each TDS receives the direct output trigger information from two VMMs.  The TDS serves as a gigabit transceiver and data serializer, aggregating all the hit data and sending the data downstream to the off-wedge trigger electronics via a 4.8 Gbps link.  The TDS trigger data is not collected or used in this test beam and only readout data collected using a scintillator trigger will be used.  The short description given here is for completeness of the documentation.

\subsection{The miniDAQ System}
\label{sec: miniDAQ}

We use the miniDAQ system developed at the University of Michigan to configure, control, and readout the front-end electronics.  The miniDAQ is a portable FPGA based DAQ system that can control and readout up to 8 front-end boards at the same time.  The miniDAQ can be directly controlled by a computer through the UDP Ethernet protocol.  Its ease of use, plus its small size made it an essential piece of equipment in the R\&D phase of the sTGC detector.  The miniDAQ was used at multiple testing sites for cosmic testing of the detector, in multiple test beams before the final FELIX DAQ system~\cite{FELIX} was serviceable, and during the final integration and commissioning of the detector as a quick secondary system to debug issues.

\begin{figure}[h!]
\centering
\includegraphics[scale=0.7]{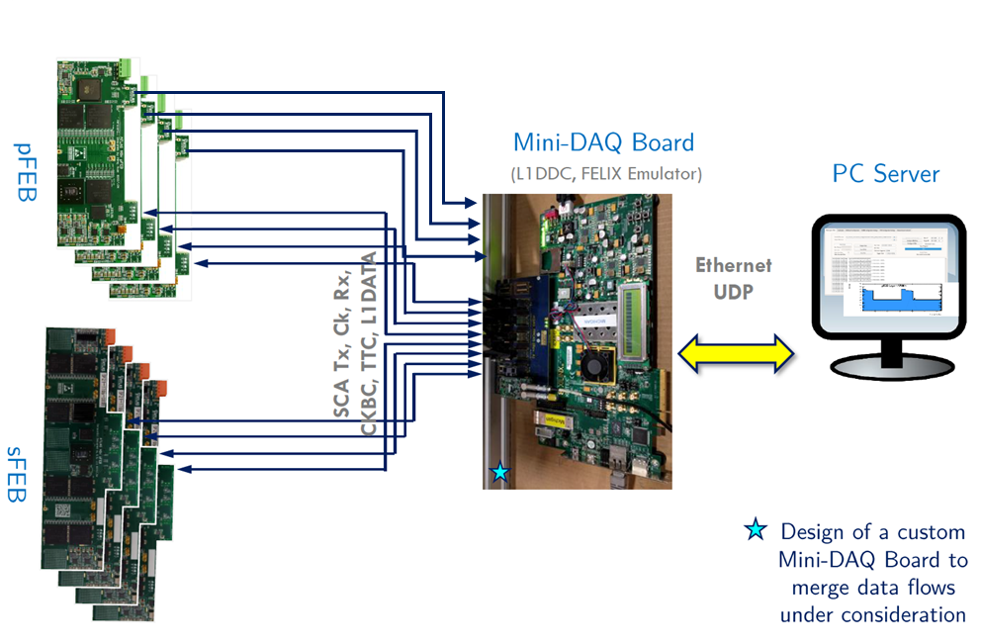}
\caption{\protect The miniDAQ system used to control and readout the front-end boards. }
\label{fig: miniDAQ}
\end{figure}

The miniDAQ hardware is composed of a KC705 Xilinx Kintex 7 FPGA evaluation board with a FSM to miniSAS adapter board to allow connections to the front-end boards.  Another small custom PCB is used to convert the NIM signal from the scintillator coincidence trigger into the LVDS level that the FPGA recognizes.  A diagram of the setup can be seen in figure \ref{fig: miniDAQ}.

The miniDAQ's functionality includes sending configurations to the FEB ASICs, listening for proper slow control replies,  monitoring FEB status such as temperature, and calibrating the FEB ASICs' parameters such as baselines, thresholds etc.  The miniDAQ also serves to distribute the global 40 MHz clock and the synchronous timing trigger control (TTC) signal and to ensure the proper alignment between the two.  It can also receive external triggers from a scintillator coincidence in order to trigger on muons or it can generate an internal trigger to readout the internally generated test pulse signal.  Finally, the miniDAQ serves as a back-end data acquisition system with the ability to buffer data from the front-end boards and pass that data to the computer via an Ethernet link.

\section{sTGC Muon Test Beam in a High Background Environment in 2018}
\label{sec: TB October}

The October 2018 sTGC test beam was the first direct measurement of sTGC efficiency with near final front-end electronics in a high background environment.  We were able to demonstrate that the sTGC efficiency vs background rate distribution is an exponential decay , which depended on a single parameter: the "effective" dead-time, which includes both front-end electronics dead time and detector signal shape. This result not only gave a strong prediction of expected sTGC efficiency for any given background rate.  It also demonstrated that the minimization of effective dead-time is key for efficient running at high rate.  The minimization of the dead-time using different methods is the focus of a later detector test covered in section \ref{sec: GIF++ analysis}.

\subsection{Experimental Setup}

We measured the hit efficiency of a single sTGC layer in a high rate environment using near final front-end boards including all final ASICs in October 2018 at the  Gamma Irradiation Facility (GIF++) at CERN Super Proton Synchrotron (SPS) North Area~\cite{GIF++}.  GIF++ provides a strong 13 TBq $^{137}$Cs source.  The $^{137}$Cs source provides an even irradiation of photons in the hundreds of keV range which mimics the effect of long-lived neutral particles from the cavern background.  A wide combinations of filters can be applied to control the incident rate of photons onto our detectors.

The CERN SPS simultaneously provides a beam of energetic muons (80 to 150~GeV) which is incident on the central part of the sTGC detector.  We trigger on the muons using a series of paddle scintillators that are in-line with the beam.  The scintillators are shielded from high energy photons by enclosing them in lead cases, plus two scintillators are placed in the beam line outside the concrete bunker, where no photons will be present. The scintillators are set up such that only muons which traverse a single detector readout pad are triggered on. A coincidence between all the scintillators marks the presence of a real muon that traversed the detector.

We setup a QL1 sTGC quadruplet equipped with a single pad front-end board described in section \ref{sec: intro_FEB}.  The miniDAQ system described in section \ref{sec: miniDAQ}is used to configure the front-end, distribute clock and the scintillator trigger signal and acquire data.

A small TGC detector is also placed in front of the sTGC chambers as a simple reference counter. It is read out by the fast ASD readout \cite{SONY_ASD}, which only digitizes a yes or no signal with little dead-time to get an in-situ reading on the rate of background photons incident on the sTGC detector.  The TGC has the same gas gap, wire and cathode construction as the sTGC and has the same detector response to the high energy photons as the full-sized sTGC. The gain of the TGC detector was set to give 8 pC induced charge per photon~\cite{Photon_charge}. 

\begin{figure}[h!]
\centering
 \begin{subfigure}[b]{0.52\textwidth}
\includegraphics[width=\textwidth]{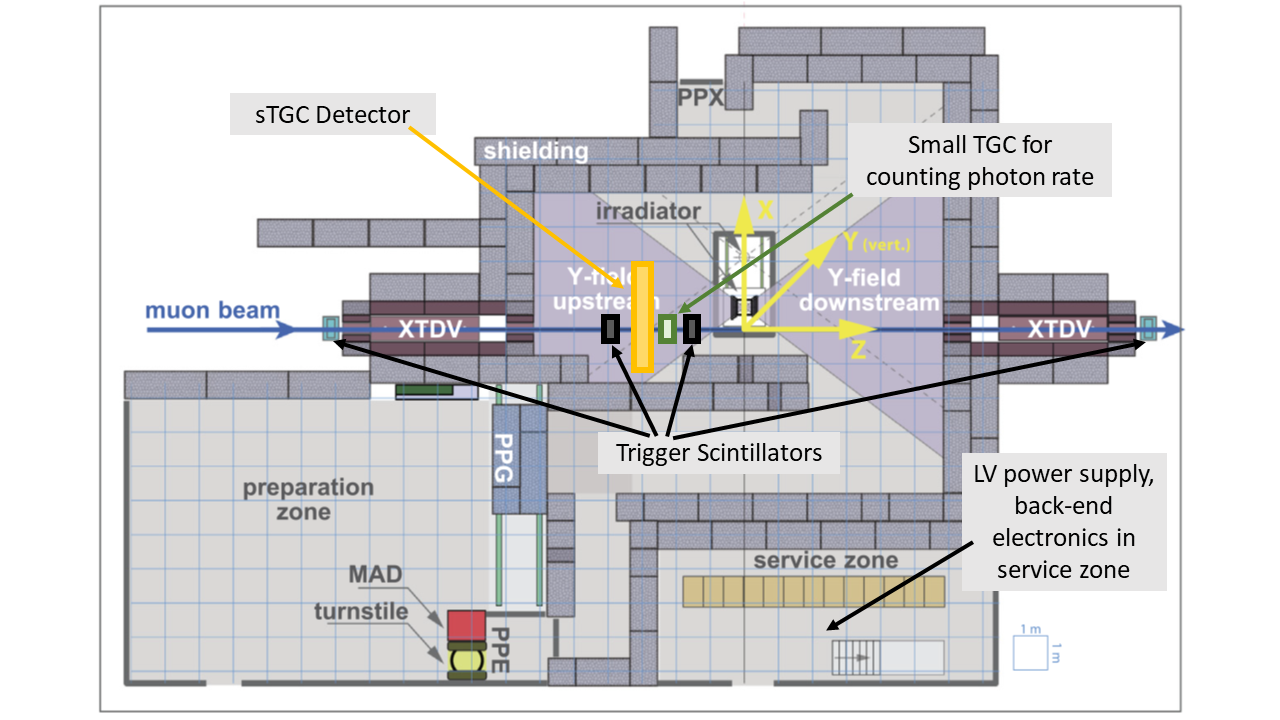}
\caption{}
\label{subfig: GIF_Oct2018_setup}
 \end{subfigure}
\qquad
 \begin{subfigure}[b]{0.4\textwidth}
\includegraphics[width=\textwidth]{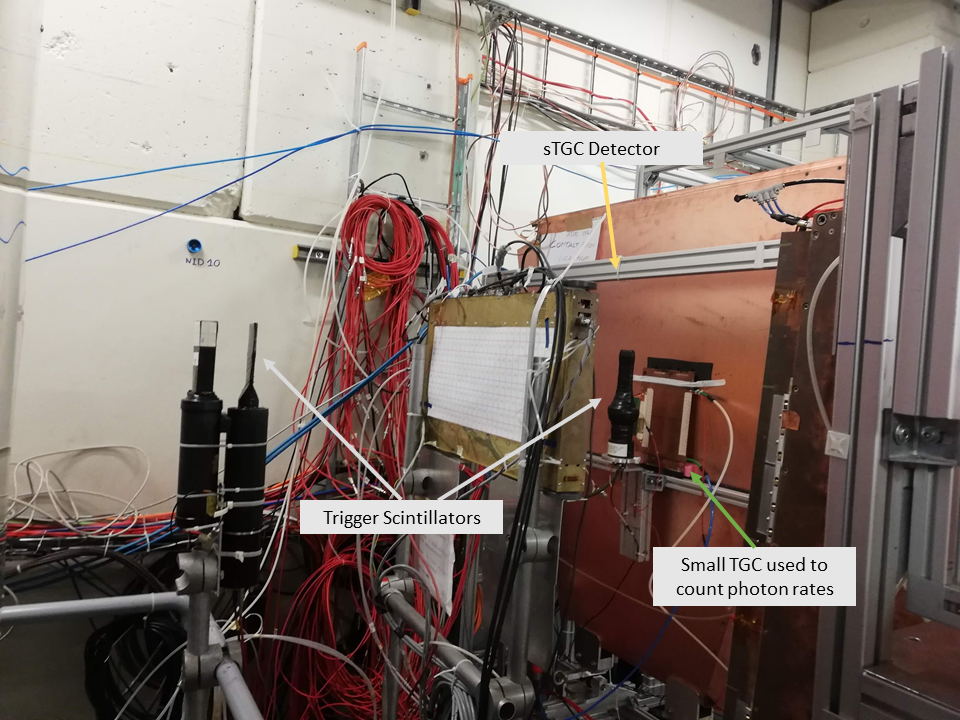}
\caption{}
\label{subfig: GIF_Oct2018_photo}
\end{subfigure}
\caption{ \ref{subfig: GIF_Oct2018_setup}: location of the detector in the GIF++ facility (taken from \cite{GIF++}). \ref{subfig: GIF_Oct2018_photo}: setup of the TGC and sTGC detectors and scintillators.}
\label{fig: GIF_setup}
\end{figure}

A photo of the location in the GIF++ beam facility and the detector setup can be seen in figure~\ref{fig: GIF_setup}.

\subsection{Method used to Measure the Detector Hit Efficiency}
\label{subsec: DetectorHitEfficiency}

\begin{figure}[h!]
\centering
\begin{subfigure}[b]{0.31\textwidth}
\includegraphics[width=\textwidth]{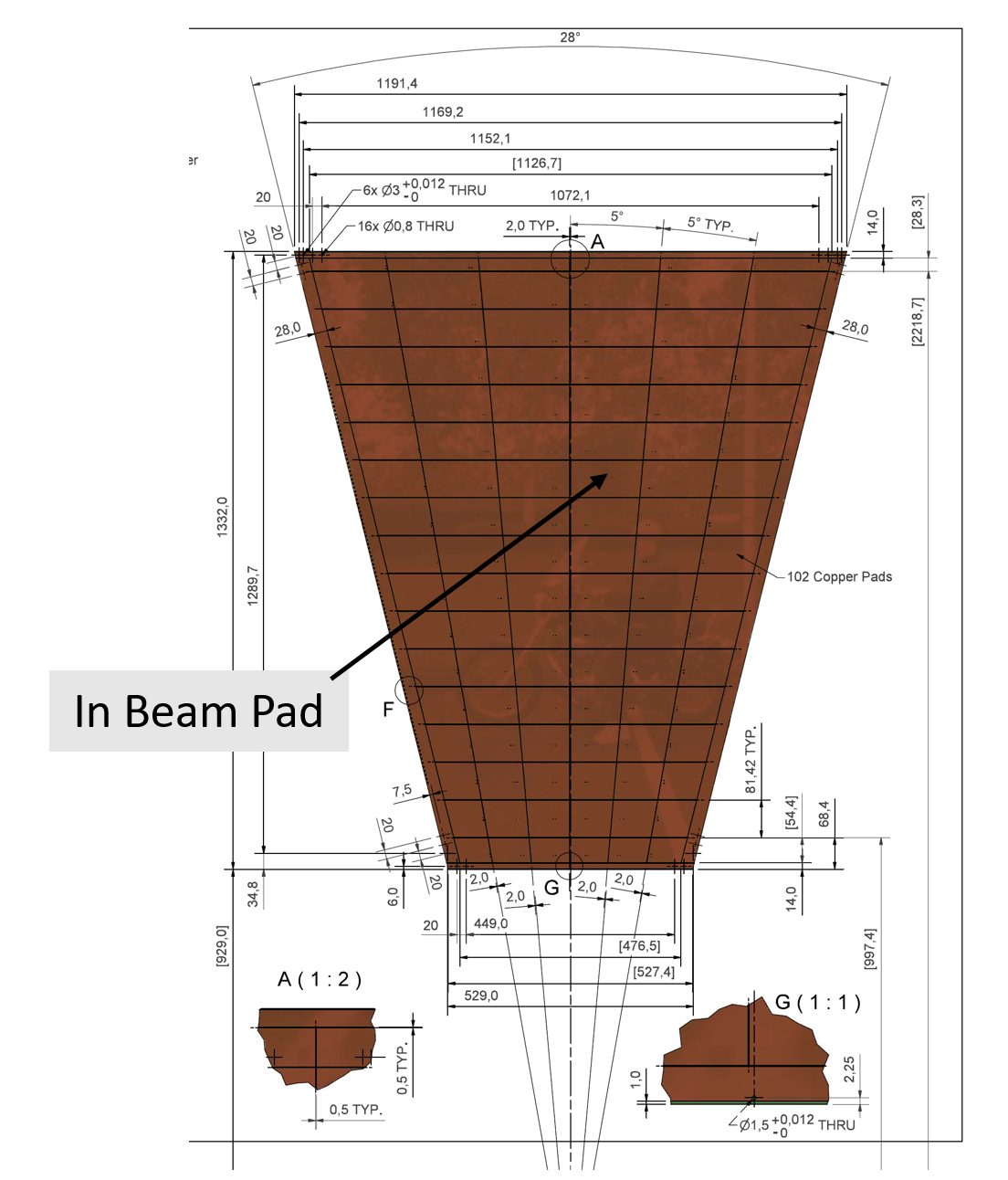}
\caption{}
\label{subfig: Det_Layer_Diagram}
\end{subfigure}
\qquad
\begin{subfigure}[b]{0.63\textwidth}
\includegraphics[width=\textwidth]{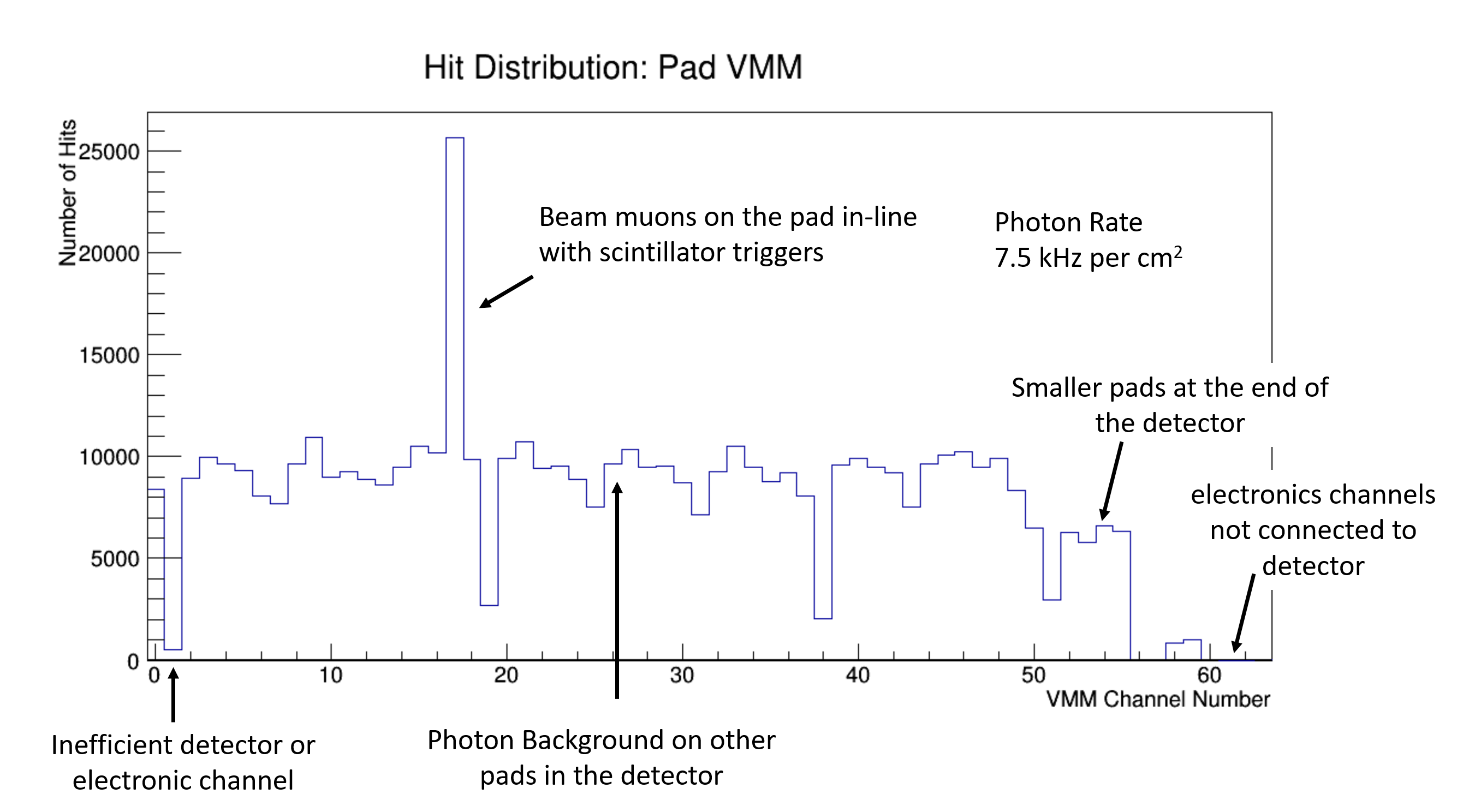}
\caption{}
\label{subfig: Det_Hit_Distribution}
\end{subfigure}
\caption{\ref{subfig: Det_Layer_Diagram} Location of the pad which is in-line with the scintillator triggers. Triggered muons traverse this pad. \ref{subfig: Det_Hit_Distribution} The hit distribution for all electronic channels. The pad channel where the muon passes has the most hits. Background photons hit all pads including the pad which the muon passes through but also all other pads.  Some inefficient detector channels can also be seen. These channels are not used in the analysis.}
\label{fig: GIF_Oct2028_pad_hit_distribution}
\end{figure}

\begin{figure}[h!]
\centering
\includegraphics[width=0.7\textwidth]{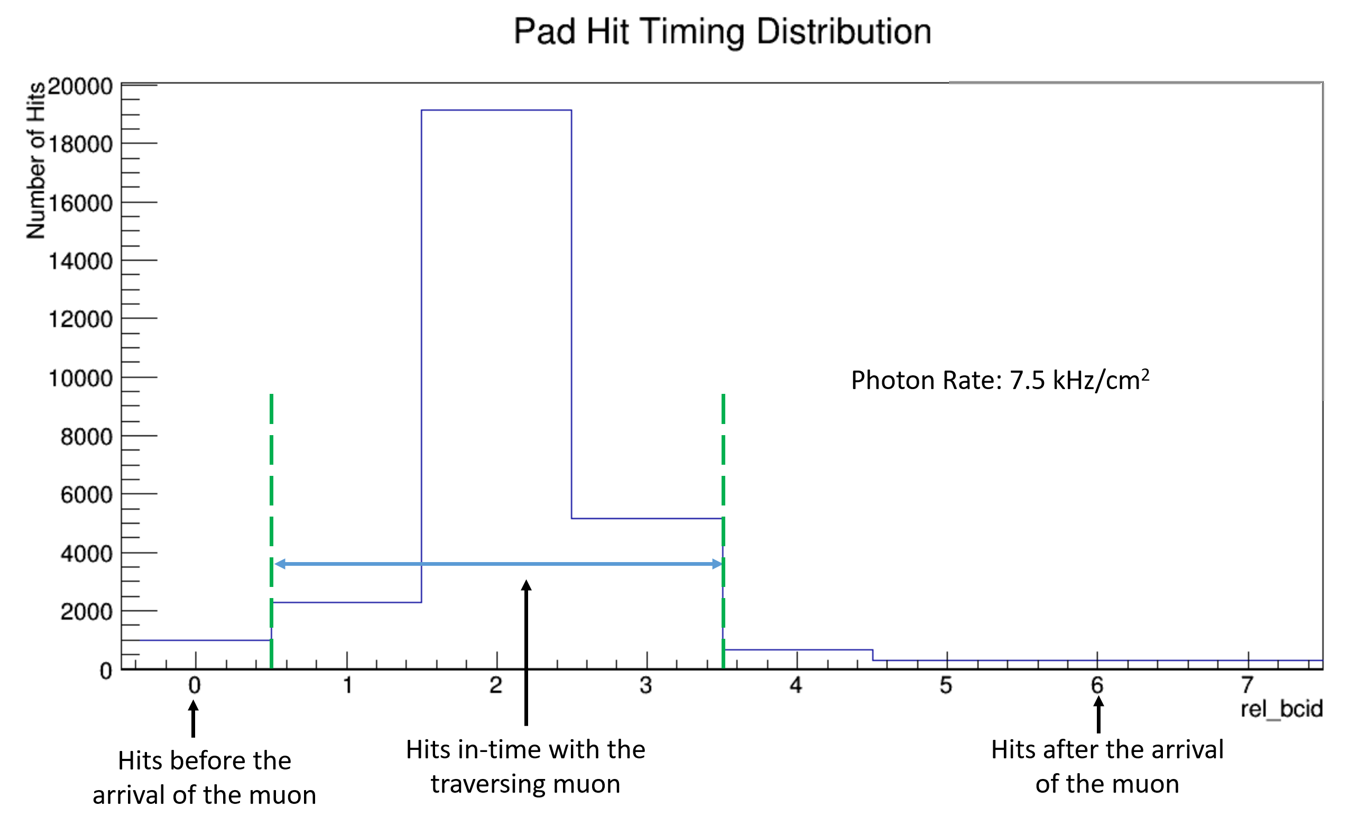}
\caption{\protect VMM hit time distribution from the pad channel in-line with the scintillator trigger. The $x$-axis is plotted in relative BCID units of time corresponding to a single 40 MHz clock-cycle or 25 ns.}
\label{fig: GIF_Oct2018_pad_time_distribution}
\end{figure}

The distribution of hits versus pads are shown in figure~\ref{fig: GIF_Oct2028_pad_hit_distribution}.
The time distribution of the recorded hits for the pad which the muon traverses is shown in figure~\ref{fig: GIF_Oct2018_pad_time_distribution}. The $x$-axis shows the relative bunch crossing identification (BCID) of the hits. Each BCID is 25 ns, corresponding to one cycle of the 40 MHz global LHC clock. The BCID of a hit is the time of the hit's peak detected by the VMM. 

 The VMM settings were such that relative BCID 2 is the clock-cycle that is exactly in-time with the scintillator trigger.  All other BCIDs are defined relative to the in-time BCID.  BCID 1 is 1 clock cycle or 25ns before the in-time BCID and BCID 3 is 1 clock cycle after the in-time BCID.
 
 Due to the time resolution of both the detector and the electronics, hits coming at relative BCID 1 or 3 are also considered within the in-time window, since they most likely arise from triggered muons. Out-of-time hits from photon background can be seen in relative BCID 0 ($50 - 25$~ns before the arrival of the muon) and relative BCID 5, 6, and 7 ($75-150$~ns after the arrival of the muon).  Relative BCID 4 has a mixture of muon hits and late photon hits and is not used for analysis. 
 
It is important to note that a photon arriving before the muon is not the same as arriving after it.  This is because all hits incur effective dead-time. If the VMM digitizes the muon hit, it will not recover in time to digitize and record another photon hit within the 200 ns readout window.  Therefore the only way for a photon hit which arrives after the muon to be digitized is if the muon hit previous to the photon is inefficient.  As a result, fewer photon hits per 25~ns are recorded after the muon compared with before the muon.

 The average arrival time of muons (in unit of BCID) is calculated according to eq.~\ref{eqn: avg_muon_time} using hits recorded with relative BCID $1-3$:
 \begin{equation}
     t_{\text{avg}}^{\text{muon}} = \frac{ \sum_{k=1}^{k=3} n_{k}^{\text{hits}} \times k } {\sum_{k=1}^{k=3}n_{k}^{\text{hits}}}.
    \label{eqn: avg_muon_time}
 \end{equation}
 
 We use out-of-time hits that relative BCID 0 and 7 as estimates for the number of photon hits arriving every BCID. Then we multiply the expected rate of photons per unit time (in BCIDs) by the amount of time before and after the muon within the 3 BCID time window. Therefore, the number of photon hits in the relative BCID $1-3$ window can be calculated as the number of early photon hits in relative BCID 0 bin times the average arrival time of the muon minus 0.5 where 0.5 is the start of the relative BCID 1 bin plus the number of late photon hits in relative BCID 7 times $(3.5 - t_{\text{avg}}^{\text{muon}})$ where 3.5 is the end of the relative BCID 3 bin.  This is shown in eq.~\ref{eqn: photon_hits_in_time}:
 \begin{equation}
    n_{\text{hits}}^{\text{photon~in-time}} = n_{\text{hits}}^{\text{relative~BCID~0}} \times (t_{\text{avg}}^{\text{muon}}-0.5) + n_{\text{hits}}^{\text{relative~BCID~7}} \times (3.5 - t_{\text{avg}}^{\text{muon}}).
    \label{eqn: photon_hits_in_time}
 \end{equation}

The number of muon hits is the total number of hits in the relative BCID $1-3$ window minus the number of photon hits in the same window according to eq.~\ref{eqn: muon_hits}:
 \begin{equation}
    n_{\text{hits}}^{\text{muon}} = n_{\text{hits}}^{\text{in-time}} - n_{\text{hits}}^{\text{photon~in-time}}.
    \label{eqn: muon_hits}
 \end{equation}

The muon efficiency is then the number of muon hits divided by the total number of muon scintillator triggers fired ($n_{\text{events}}$) according to eq.~\ref{eqn: muon_efficiency}:
 \begin{equation}
    \epsilon^{\text{muon}} = \frac{n_{\text{hits}}^{\text{muon}}}{n_{\text{events}}}. 
    \label{eqn: muon_efficiency}
 \end{equation}

\subsection{Method used to Measure the Background Photon Rate}
\label{sec: bkg photon rate}

The background photon rate is measured both using the sTGC and the small TGC monitor detector.  The two methods serve as a cross-check to each other as both have limitations.

Directly measuring the photon background rate using the sTGC with VMM has the draw-back that photon hits are also subject to any inefficiency that muon hits experience, as described in section~\ref{sec: intro_VMM}. Similarly to the muon, the true photon efficiency can be recovered from the measured one by dividing the measured rate of photon hits (hits that are out-of-time and come before the muon) by the muon efficiency.  This assumes that the muon efficiency is the same as the photon efficiency, both caused by photon hits.  In fact, the inefficiency caused by muons is negligible, given their low flux with respect to the photon one.

The second method of measuring photon rate is using the small TGC monitor detector placed directly in front of the sTGC.  This gives a solid direct measurement of the incident photon rate as the fast ASD of the TGC has little dead-time \cite{SONY_ASD}.  However the measured small TGC photon hit rate cannot be directly applied to the sTGC photon hit rate.  This is because the sTGC is segmented into many pads and strips.  Therefore a single photon may generate more than one hit in the sTGC.  Hence a scale factor need to be determined for the small TGC vs sTGC hit rates.

\begin{figure}[h!]
\centering
\includegraphics[width=0.7\textwidth]{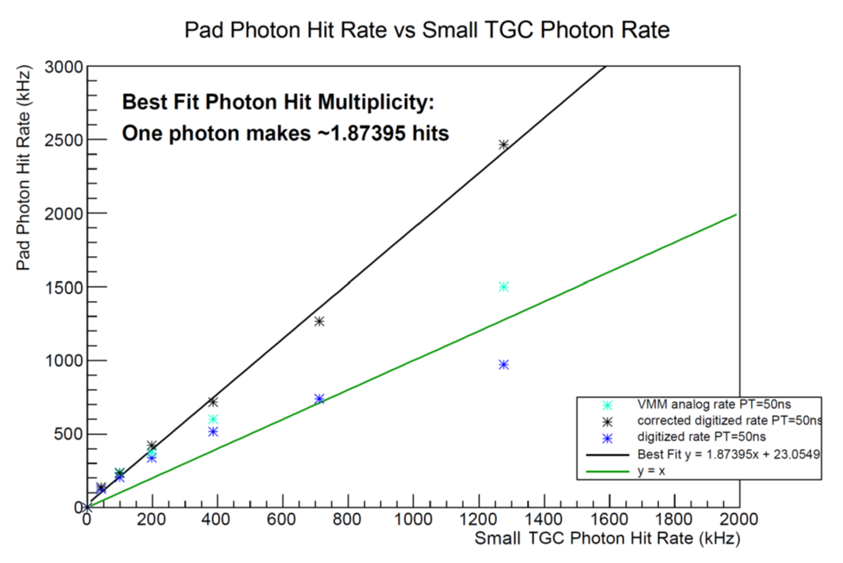}
\caption{\protect Distribution of sTGC pad photon hit rate as a function of the small TGC photon hit rate.  The dark blue dots are the raw sTGC hit rate measurements.  The black dots are the sTGC hit rates divided by the muon efficiency, therefore correcting for any detector and electronics inefficiency.  Linear relationship between the corrected sTGC photon hit rate and the TGC photon hit rate can be seen. The hit multiplicity in the sTGC pad is $\sim$1.9 estimated from the linear fit. The rate of analog output pulses after the shaper measured with an oscilloscope is also presented with cyan dots. The $y=x$ line is shown as reference.}
\label{fig: GIF_Oct2018_PhotonRate}
\end{figure}

Figure \ref{fig: GIF_Oct2018_PhotonRate} shows the distribution of the sTGC pad photon hit rate versus the TGC photon hit rate. A linear relationship can be seen between the two after correcting sTGC hits for the efficiency loss.  A clear saturation can be seen in the raw sTGC hit rate due to the loss of efficiency at high rate.  
In the same figure, measurements of the sTGC photon hit rate from the analog output of the amplifier and shaper using an oscilloscope are also shown. The oscilloscope measurement avoids the dead-time associated with ADC digitization and the reset of the digital logic after a hit but still includes the dead-time associated with the amplifier and shaper's need to dump the charge input from a hit and recover. It is more efficient than the digitized output, however it cannot recover itself quickly enough at 2.5 MHz per channel, as demonstrated by the loss of efficiency (i.e. linearity).  A 50 ns amplifier peaking time was chosen for this measurement.  Notice that the three graphs agree with one another at low photon rates up to $\sim$100 kHz. 

From the slope of the linear fit of the corrected sTGC hit rates it is possible to extract the hit multiplicity. For every photon hit on the small TGC detector, around 1.87 hits are seen in the sTGC pads.  This is because a single photon can make multiple hits in different sTGC pads.  The same phenomenon is also observed with muons in the absence of any photon background. A single muon can make more than one hit in $\sim$10\% of events. The small TGC on the other hand is not segmented into multiple electronic channels so only a single hit is possible for every muon.  Appendix \ref{sec: investigations} covers this phenomenon in more detail.  The true cause probably includes a combination of several different modes of propagation and we do not yet have a full understanding of how the signal is propagated between different pads. This "cross-talk" effect should be investigated thoroughly by means of actual collision data. Defining operation parameters to minimize it might be important for a good performance of the detector.  These additional hits are correlated in space and time with the original hit and could confuse the trigger algorithm. At the same time, the additional hits generate additional effective dead-time and cause efficiency loss at high rates.

\subsection{Deriving the Relationship Between Efficiency and Effective Dead Time in Theory} 
\label{subsec: efficiency_theory}

The photons arrival time at the detector follows  a Poisson distribution with expected rate $\lambda$. Therefore, the time \textit{t} in between two photons or a muon and a photon directly preceding the muon is given by the exponential distribution of equation~\ref{eqn: t_between_hits}:
 \begin{equation}
    P(\text{two~hits~separated~by~time~$t$}) = \lambda \times e^{-\lambda t}.
    \label{eqn: t_between_hits}
 \end{equation}

The average amount of incurred effective dead-time per hit if a single hit normally takes time \textit{D} to recover is given by eq.~\ref{eqn: avg_dead_time}:
\begin{equation}
    t^{\text{dead-time}}_{\text{avg}} = \int^D_0 t \times \lambda e^{-\lambda t} dt + \int^{\infty}_D D \times \lambda e^{-\lambda t} dt
    = \frac{1}{\lambda}( 1-e^{-D\lambda} ).
    \label{eqn: avg_dead_time}
 \end{equation}
The reason why the average amount of incurred effective dead-time per hit is not just the dead-time per hit \textit{D} is because hits from photons have a probability of occurring in the dead-time of a previous hit.  When two photon hits occur within each other's dead-time then the full dead-time of the first hit is not realized.  This lack of fully-realized dead-time is given by the first integral between 0 and time \textit{D}, when the total amount of incurred dead-time is only the time between the first hit and the second hit.  The second integral from \textit{D} to infinity represents the possibility that the second hit arrives after the first hit's dead-time and therefore the full dead-time \textit{D} is realized by the first hit.
 
The fraction of the time that the electronic channel is inactive is given by the average effective dead-time per hit $t^{\text{dead-time}}_{\text{avg}}$ multiplied by the hit rate $\lambda$.  The efficiency is simple one minus the fraction of the time the electronics is in dead-time:
 \begin{equation}
    \epsilon = 1 - \lambda \times t^{\text{dead-time}}_{\text{avg}}
    = e^{-D\lambda}.
    \label{eqn: efficiency_exponential}
 \end{equation}
This calculation is valid under the assumption that a second hit will incur additional effective dead-time even if it occurs within the dead-time of the first hit.  This can be seen in its asymptotic behavior at infinite rate where the measured rate, or $\lambda \times \epsilon$, would approach zero. Further discussion on difference assumptions on whether hits within effective dead-time also generate dead-time are explored in the appendix \ref{app: efficiencyVsRateTheory}.

This calculation also assumes that the loss of efficiency is only due to effective dead-time including both front-end electronics dead time and detector signal shape, with no other contributing factors.  For example, if the cyclical FIFO began to fill up at some rates and hits are lost in the digital readout chain then we expect some emergent behavior at high rate that differs from the simple equations above. We did not observe any such deviation at high rates. The lack of any deviation at high rates can be seen as evidence that the efficiency loss due to front-end dead-time is the primary cause of efficiency loss at high rates.

The effect of detector inefficiency at high particle flux, due for example to large currents causing a voltage drop on the resistive layer, is also considered negligible here. This assumption is justified by the spectra shapes, showing a good signal to noise separation in all irradiation conditions. This is confirmed by a separate study with external pulses injected into the electronics which emulate the charge deposited by photons at high rates~\cite{DetEmulation}. 

We chose to fit our data using the exponential decay function shown in eq.~\ref{eqn: efficiency_exponential_fitted} and extract the effective dead-time $D$.  There are small region of dead area which are inefficient in the detector such as plastic wire supports and buttons which hold in place the anode wires.  The $\epsilon_0$ parameter is introduced to account for the fact that the test chamber we used is not fully efficient even when no background hits are present:
 \begin{equation}
    \epsilon = e^{-D\lambda} - \epsilon_0. 
    \label{eqn: efficiency_exponential_fitted}
 \end{equation}

\subsection{Results: The Muon Efficiency vs Rate and the Measured Effective Dead-Time} 
\label{sec: meas eff vs dead time}

\begin{figure}[h!]
\centering
\includegraphics[scale=0.55]{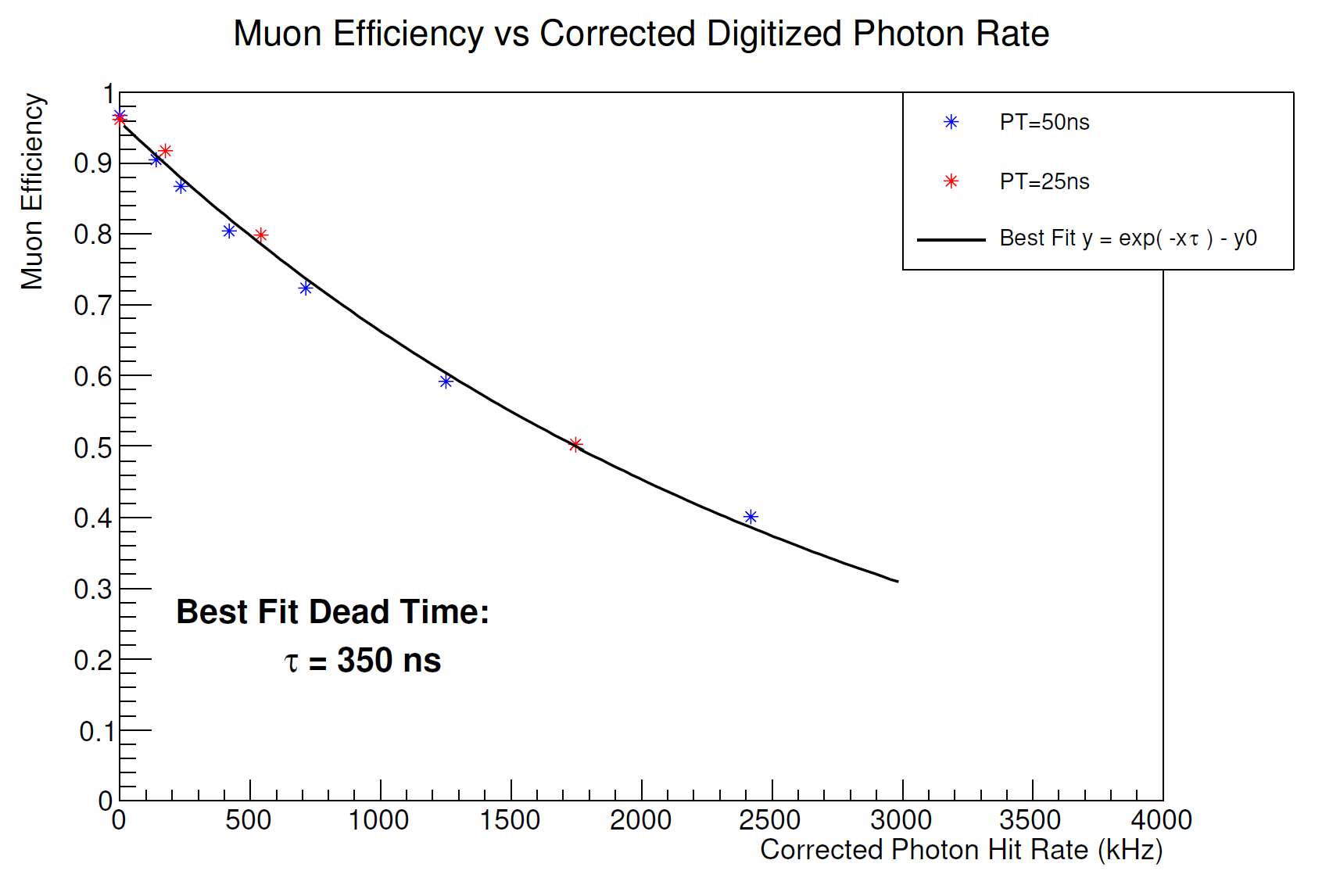}
\caption{\protect Distribution of the measured efficiency vs the background photon hit rate with 10-bit ADC mode and $C_p=100$ pF. Inefficiencies have been taken into account in the calculation of background photon hit rate. }
\label{fig: GIF_Oct2018_EfficiencyCurve}
\end{figure}

The muon efficiency versus background hit rate for a sTGC pad is shown in figure~\ref{fig: GIF_Oct2018_EfficiencyCurve} with 10-bit ADC mode and the default 100 pF $C_p$. The hit rate is corrected for inefficiency according to the procedure described in section \ref{sec: bkg photon rate}. The relationship can be well fitted by the exponential decay function in eq.~\ref{eqn: efficiency_exponential_fitted}. The value of $\epsilon_0$ is about 4\% and it represents the intrinsic inefficiency of the detector. The best fit value for the effective dead-time \textit{D} is 350~ns. No sharp deviation from the exponential fit is observed up to a rate of 2.5~MHz suggesting that electronics-related dead-time is the only measurable cause of efficiency loss up to that rate.

Both 25 ns and 50 ns peaking time show similar efficiency versus background hit rate. The effective dead-time for peaking times are similar in this case may be because the dead-time is most likely dominated by the 10-bit ADC conversion time.  This shows that for actual operation at high rates (15\,kHz/cm$^2$ with a 64 cm$^2$ pad equals to 1 MHz per channel in pads closest to the beam line), the efficiency loss may be significant when operating in the slow conversion time of the 10-bit ADC.


\section{Attempts at Minimizing Electronics Dead-Time in 2019}
\label{sec: GIF++ analysis}

Different methods of decreasing the dead-time were explored in March 2019 in the GIF++ facility. 
The electronics' dead-time can be decreased in two independent ways.  One method involves finding a better operating configuration for the VMM amplifier.  For example, resetting the electronics channel after finishing the fast 6-bit ADC conversion instead of waiting for the 10-bit ADC to finish decreases the dead-time after a hit. 
The second method involves optimizing the input pi-network charge filter parameter to limit the amount of input charge physically entering the detector channel.  Decreasing the input charge also decreases the time the analog pulse needs to recover to baseline after a hit.


\subsection{Experimental Setup Testing Different Input Filter Circuit Components}

The experimental setup was similar to the October 2018 GIF test beam. Two sTGC quadruplets were placed inside the GIF bunker with front-end boards using the final ASICs.  The readout was done using the miniDAQ system and an oscilloscope was used to monitor the analog output for debugging purposes.

We modified the values of the main components in the pi-network circuits on the pFEBs and pull-up resistors on the sFEBs.  A regular pattern of 100 pF, 150 pF, 200 pF, 300 pF, and 470 pF $C_p$ was installed on the pFEB channels. Likewise a regular pattern of 200 k$\Omega$, 300 k$\Omega$, 500 k$\Omega$, and 1 M$\Omega$ pull-up resistors was installed on the sFEB channels in order to test the performance of the different capacitance and resistances. A smaller resistance corresponds to more input current, which partially cancels out the input charge from the long ion tail at high rates. 

\begin{figure}[h!]
\centering
\begin{subfigure}[b]{0.4\textwidth}
\includegraphics[width=\textwidth]{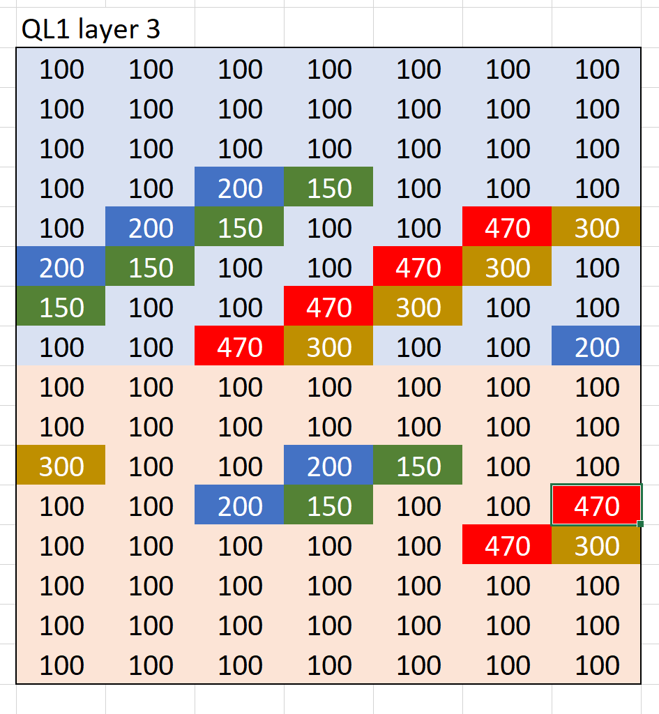}
\caption{}
\label{subfig: QL1CPiMapping}
\end{subfigure}
\qquad
\begin{subfigure}[b]{0.4\textwidth}
\includegraphics[width=\textwidth]{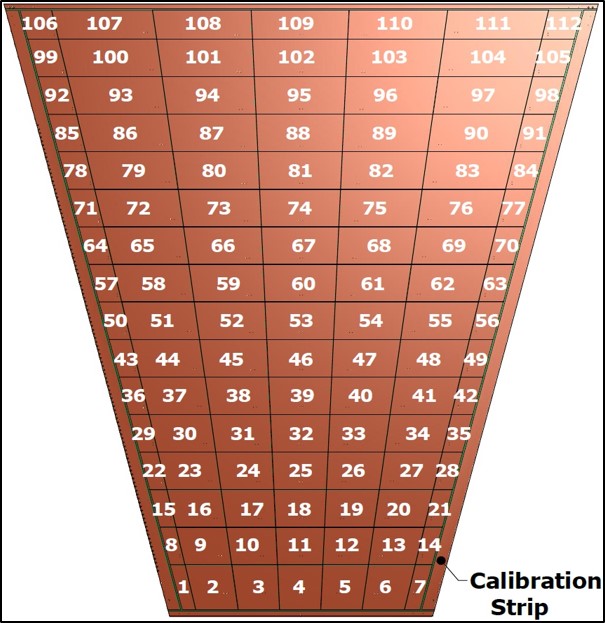}
\caption{}
\label{subfig: QL1PadMapping}
\end{subfigure}
\caption{\label{fig: QL1Mapping} A diagram of the pi-network capacitance for different QL1 pads is shown in \ref{subfig: QL1CPiMapping}. A physical-to-scale diagram of the QL1 pads is shown in \ref{subfig: QL1PadMapping}.}
\end{figure}

\begin{figure}[h!]
\centering
\begin{subfigure}[b]{0.25\textwidth}
\includegraphics[width=\textwidth]{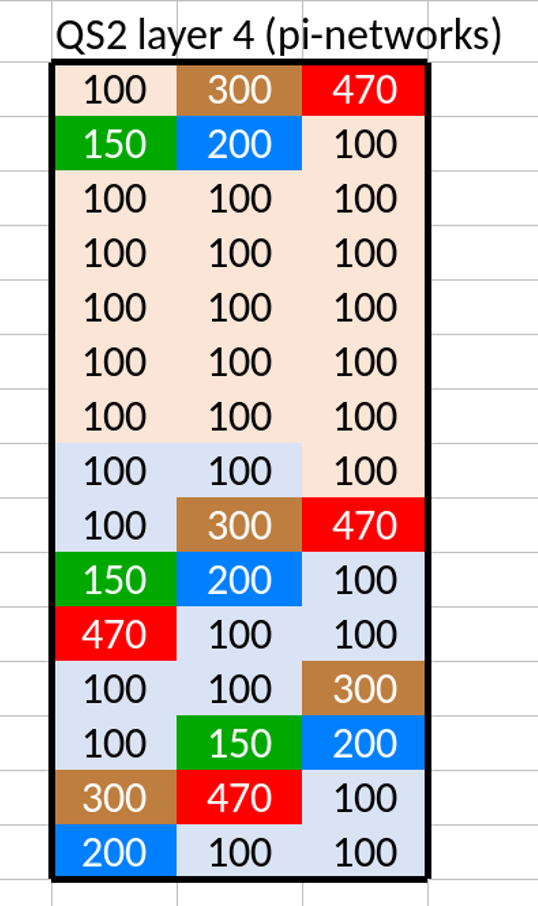}
\caption{}
\label{subfig: QS2CPiMapping}
\end{subfigure}
\qquad
\begin{subfigure}[b]{0.4\textwidth}
\includegraphics[width=\textwidth]{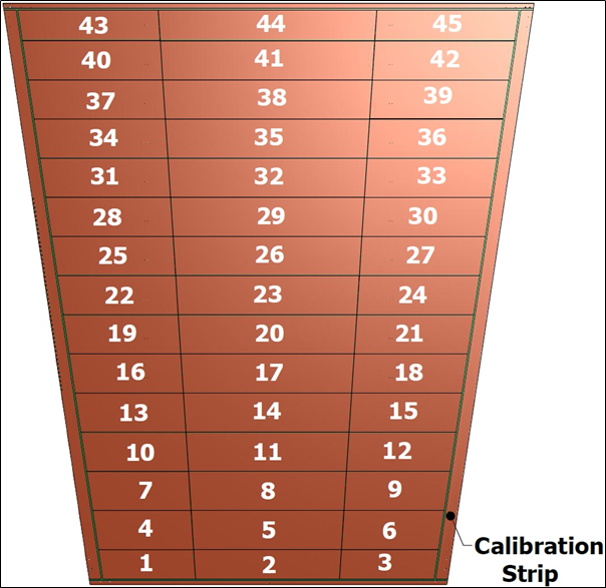}
\caption{}
\label{subfig: QS2PadMapping}
\end{subfigure}
\caption{\label{fig: QS2Mapping} A diagram of the pi network capacitance for different QS2 pads is shown in \ref{subfig: QS2CPiMapping}. A physical-to-scale diagram of the QS2 pads is shown in \ref{subfig: QS2CPiMapping}.}
\end{figure}

The sTGC's design is such that the pad size increases with respect to the radial distance from the interaction point. Moreover, three different quadruplet types are named Q1, Q2, and Q3 according to their distance from the interaction point.   The pad area is directly proportional to the pad's capacitance, and the pad's capacitance is inversely proportional to the attenuation factor of the pi-network.  Therefore, a different pi-network capacitor must be chosen for each Q1, Q2, and Q3 quadruplet type to have roughly the same attenuation factor. In this work we tested two quadruplets, a QL1 and a QS2 module, belonging to Q1 and Q2 type, respectively. There were the only available chamber types at the time of the study.  The pi-network capacitor map and pad segmentation diagram for QL1 and QS2 are shown in figures~\ref{fig: QL1Mapping} and~\ref{fig: QS2Mapping}, respectively.
To shorten the dead time, all the measurements of this optimization study were taken with 6-bit ADC mode.

\subsection{Calibration of Channel Thresholds for Different Input Filters}

The effective capacitance ($C_{\text{eff}}$) that the amplifier sees is given by $1/f$ where $f$ is the attenuation factor defined in eq.~\ref{eqn: pi-attenuation} because the pi-network capacitor is placed in between the amplifier input and the detector channel:
In practice, the effective capacitance is approximated by that of $C_p$, which is chosen to be a factor of $3-7$ smaller than $C_{\text{detector}}$ in order to meaningfully attenuate the detector signal.  

The amplifier noise increases linearly with the capacitative load. 
Other noise sources such as irradiated noise from the DC-DC converter are also significant contributors. 

The higher the $C_p$ the lower the attenuation to both detector signal and noise, therefore channels with higher $C_p$ have higher noise and need higher thresholds.  We therefore set the threshold that corresponds to a noise rate of approximately 100 Hz.  In practice, 470 pF channels have a threshold of 60 mV above baseline. 300 pF channels has a threshold of 40 mV above baseline. 200 pF channels
has a threshold of 30 mV above baseline. 150 pF channels has a threshold of 25 mV above baseline and 100 pF channels have a threshold of 20-25 mV above baseline depending on the channel.

Changing the different pull-up resistor for different strip channels changes the parallel resistance noise for strips.  The parallel resistance noise is found to be negligible for all pull-up resistance values tested when added in quadrature with other contributions such as the input capacitance contribution.  We observed that regardless of the resistance value, the noise level is the same.  Thresholds for strips are set to 11 mV above baseline which limit the noise rate to $\sim$100 Hz for all channels regardless of the resistor value

\subsection{Front-End ASIC Configuration which Minimize the Dead-Time}

The VMM has different configurations corresponding to different channel reset procedures as described in section \ref{sec: intro_VMM}.  Specifically, we aim to minimize the dead-time by setting the channel to reset at the 6-bit ADC completion mode. In this mode, the VMM will not wait for the slow 10-bit ADC to complete. Instead, it resets the channel as soon as the fast 6-bit ADC finishes or when the analog channel is again below the threshold, whichever is slower.

We also use 25 ns peaking time instead of 50 ns for this measurement as the lower integration time decreases the rise time of the signal pulse. This in turn decreases the total time that the pulse stays above the threshold. Both of these settings are designed to minimize the dead-time incurred both in the digitization time of the ADC and the analog time to return to the baseline.  There is one additional dimension that we were not able to explore and that is raising the threshold further above the baseline.  We did not know the beginning of the muon landau for each different detector pad and $C_p$ combination and instead set the threshold according to noise.  In practice, setting the highest threshold possible without cutting into the muon landau offers two benefits.  A higher threshold means that the analog pulse stays above the threshold for a shorter amount of time and so the channel reset procedure begins earlier.  A higher threshold also means fewer background hits cross the threshold in the first place and therefore results in a lower rate per channel for the same background flux.  A lower hit rate translates to lower accumulated dead-time and a higher efficiency.  These potential improvements to performance will be optimized in future measurements.

\subsection{Method used to Measure Electronics Efficiency}
No muon beam was available during the March 2019 measurement in the GIF++ facility.  In addition, it would be impractical to measure so many different pad and strip channels one by one using a muon beam as each pad/strip must be targeted individually.
Instead, background photons irradiated the entire detector and test pulses were sent to all electronics channels.  The electronics would not be able to digitize the test pulse if it was  immediately preceded by a photon hit since  the channel was still in dead-time.  Therefore the efficiency of all electronics channels can be measured simultaneously instead of having a beam of muons which targets a single pad.  This will not capture the detector effect of the loss of gain at high rates for example but such effects were already demonstrated to be a sub-dominant effect in the October 2018 test beam in terms of efficiency loss (see section \ref{sec: meas eff vs dead time}). 
The efficiency is calculated in a way analogous to what described in section \ref{subsec: DetectorHitEfficiency} for muons.


\subsection{Results: Maximizing sTGC Pad Efficiency and Optimization of the Pi-Network Filter}

\begin{figure}[h!]
\centering
\begin{subfigure}[b]{0.45\textwidth}
\includegraphics[width=\textwidth]{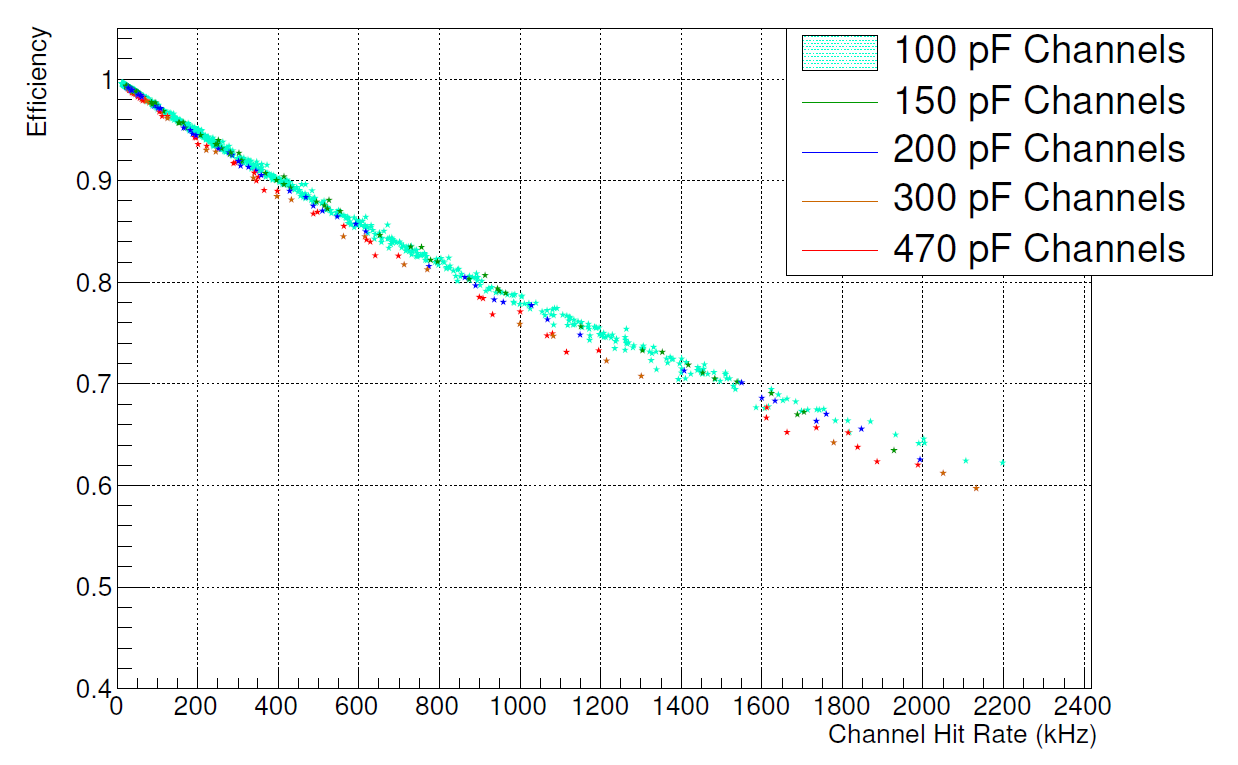}
\caption{}
\label{subfig: QL1_pad_eff}
\end{subfigure}
\qquad
\begin{subfigure}[b]{0.45\textwidth}
\includegraphics[width=\textwidth]{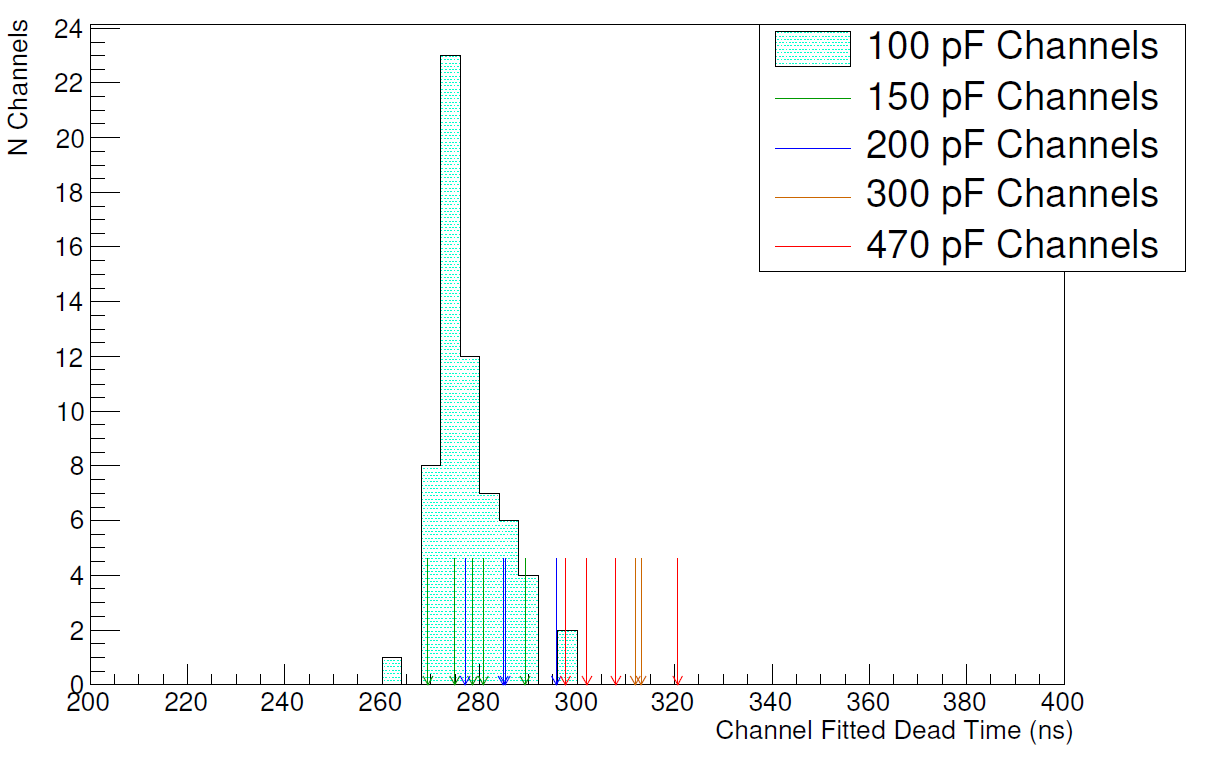}
\caption{}
\label{subfig: QL1_pad_deadtime}
\end{subfigure}
\caption{\label{fig: QL1_pad_efficiency_TP} \ref{subfig: QL1_pad_eff}: Electronics efficiency as a function of the photon hit rate for the QL1 pads with different pi-network $C_p$ values using 6-bit ADC mode. \ref{subfig: QL1_pad_deadtime}: Fitted dead-time for different $C_p$ values. }
\end{figure}

\begin{figure}[h!]
\centering
\begin{subfigure}[b]{0.45\textwidth}
\includegraphics[width=\textwidth]{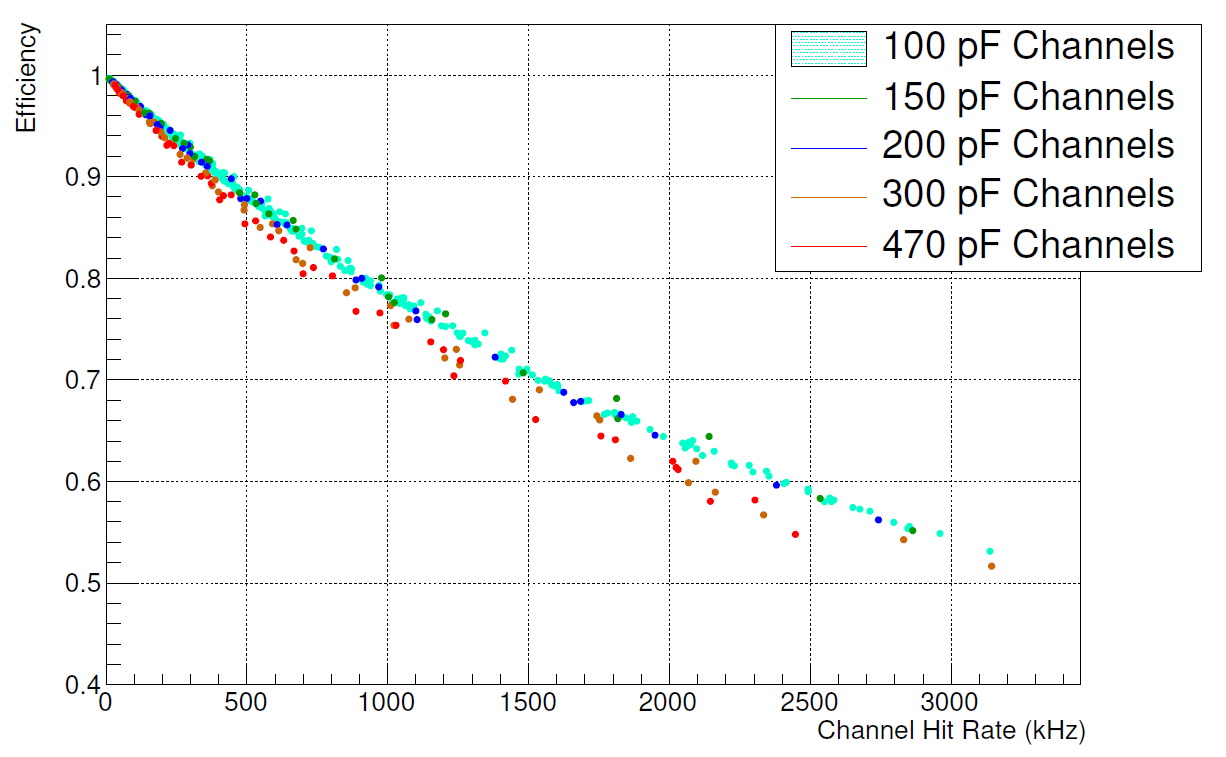}
\caption{}
\label{subfig: QS2_pad_eff}
\end{subfigure}
\qquad
\begin{subfigure}[b]{0.45\textwidth}
\includegraphics[width=\textwidth]{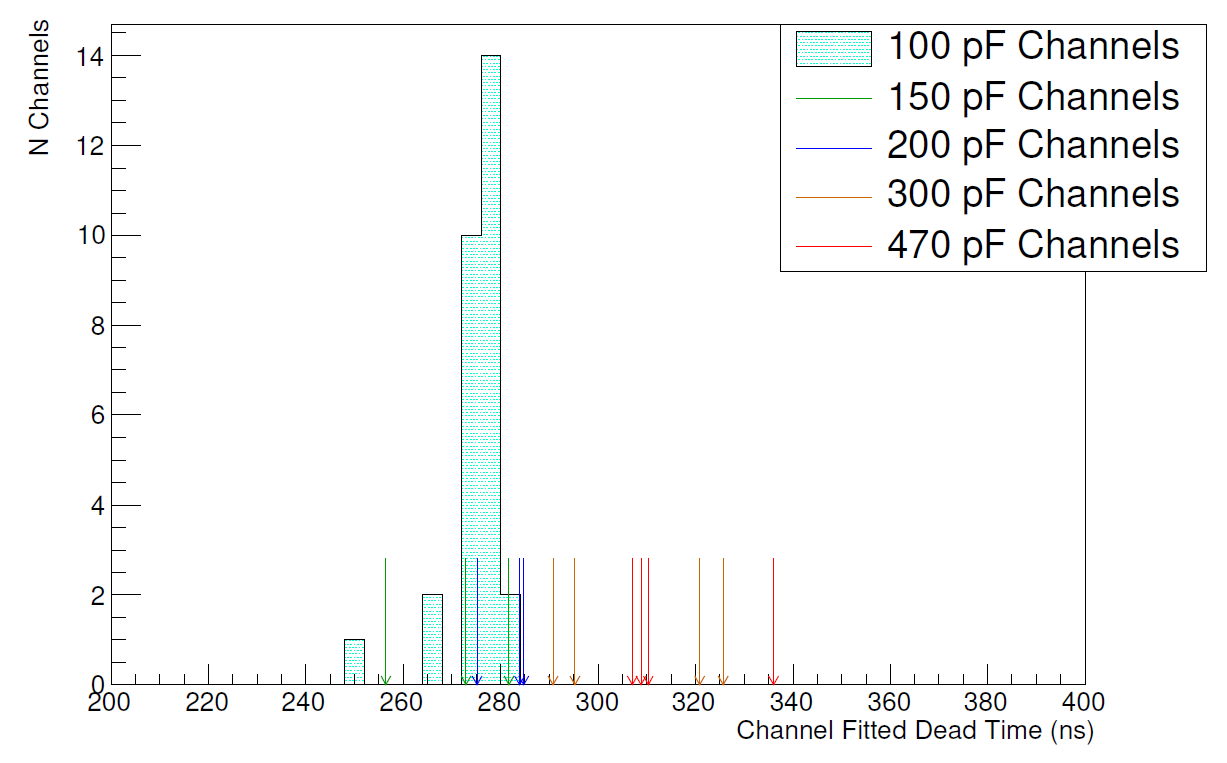}
\caption{}
\label{subfig: QS2_pad_deadtime}
\end{subfigure}
\caption{\label{fig: QS2_pad_efficiency_TP} \ref{subfig: QS2_pad_eff}: Electronics efficiency as a function of the photon hit rate for the QS2 pads with different pi-network $C_p$ values using 6-bit ADC mode. \ref{subfig: QS2_pad_deadtime}: Fitted dead-time for different $C_p$ values.}
\end{figure}

The measured electronics efficiency as a function of the photon hit rate and the fitted dead-time for the QL1 and QS2 pads are shown in figures~\ref{fig: QL1_pad_efficiency_TP} and~\ref{fig: QS2_pad_efficiency_TP}. The dead-time increases for higher $C_p$ values and therefore lower attenuation factors.  The increase is only noticeable for $C_p \ge 300$\,pF compared with $C_p = 100$\,pF.

We know from the test beam in October 2018 that a $C_p$ value of 100\,pF was able to efficiently detect muons for QL1 detectors.  Therefore we chose 200\,pF as the optimal value as it is the largest $C_p$ (least attenuation) for the Q1 detector which does not have an observable increase in the dead-time.  The attenuation in this configuration is around $1/5$. These results are also confirmed by a separate study using a pulse injector that emulates the charge input into the detector due to photons~\cite{DetEmulation}. 

For Q2 detectors we picked 330\,pF as the optimal $C_p$ value.  The value is chosen to maintain roughly the same $1/5$ attenuation factor as the QL1.  There were no muon beam or cosmic muon data at the time for the QS2 detector measured with different $C_p$.  However, QL1 detectors could measure muons with a $1/10$ attenuation factor.  This observed margin in QL1 means that a Q2 detector should be able to measure muon signals with significant margins.  

For Q3 detectors we picked 470\,pF as the optimal $C_p$ value as this corresponds to also a $1/5$ attenuation factor for the larger Q3 pads which can be as large as 3 nF.

The final pad front-end boards used these optimized values for their pi-networks \cite{sTGCFEB}. The ability to detect muons with greater than 95\% efficiency using these $C_p$ values were later confirmed in cosmic muon data using the final detector and front-end board measurements in 2020 and 2021.  The equalized attenuation factor is also confirmed as all pads have similar Most Probable Values for the Landau distribution. 

\subsection{Results: Maximizing sTGC Strip Efficiency and Optimization of the Pull-Up Resistor}

\begin{figure}[h!]
\centering
\begin{subfigure}[b]{0.45\textwidth}
\includegraphics[width=\textwidth]{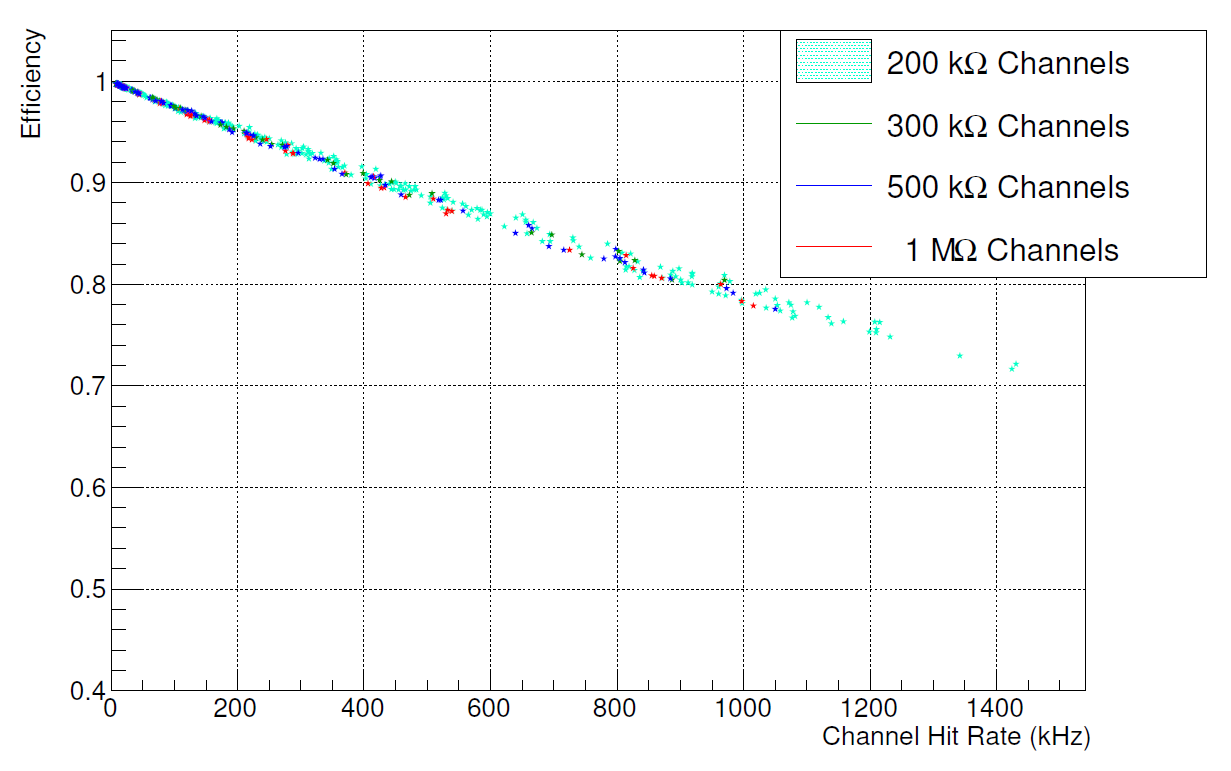}
\caption{}
\label{subfig: QL1_strip_eff}
\end{subfigure}
\qquad
\begin{subfigure}[b]{0.45\textwidth}
\includegraphics[width=\textwidth]{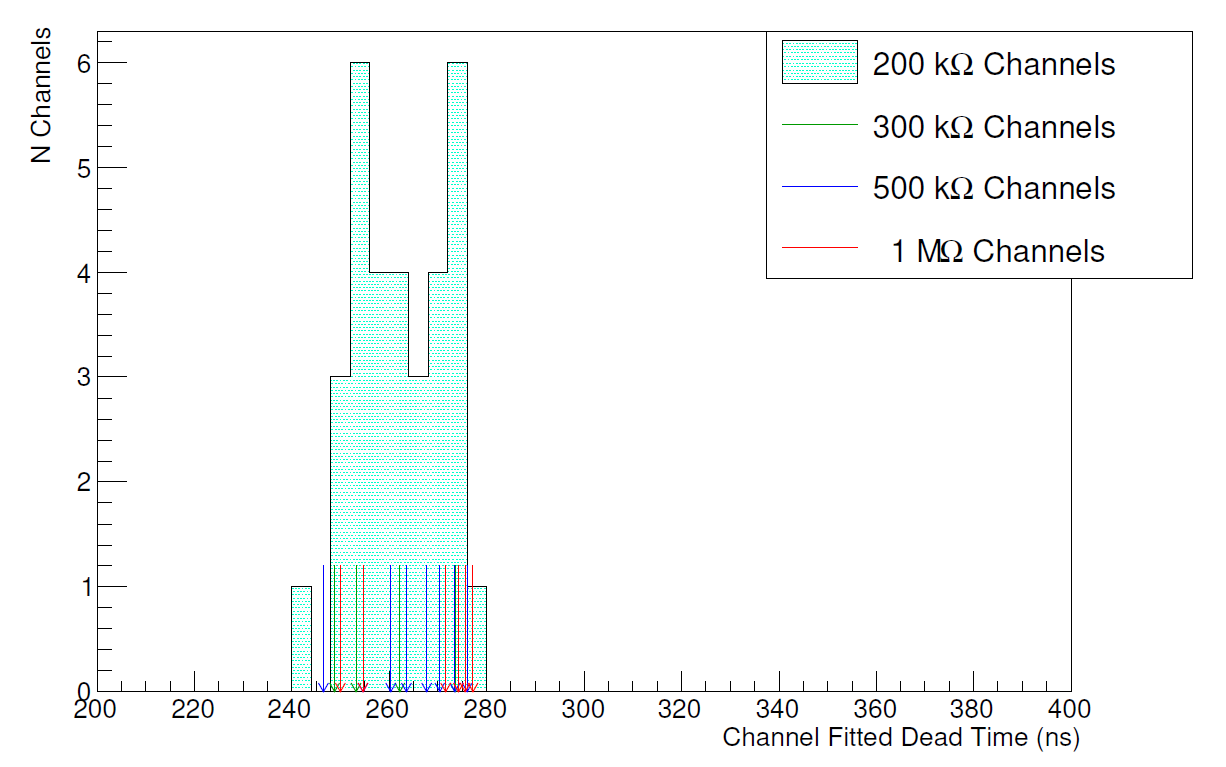}
\caption{}
\label{subfig: QL1_strip_deadtime}
\end{subfigure}
\caption{\label{fig: QL1_strip_efficiency_TP} \ref{subfig: QL1_strip_eff}: Electronics efficiency as a function of the photon hit rate for the QL1 strips with different pull-up resistor values using 6-bit ADC mode. \ref{subfig: QL1_strip_deadtime}: Fitted dead-time for different pull-up values for the QL1 strips for the 6-bit ADC mode.}
\end{figure}

\begin{figure}[h!]
\centering
\begin{subfigure}[b]{0.45\textwidth}
\includegraphics[width=\textwidth]{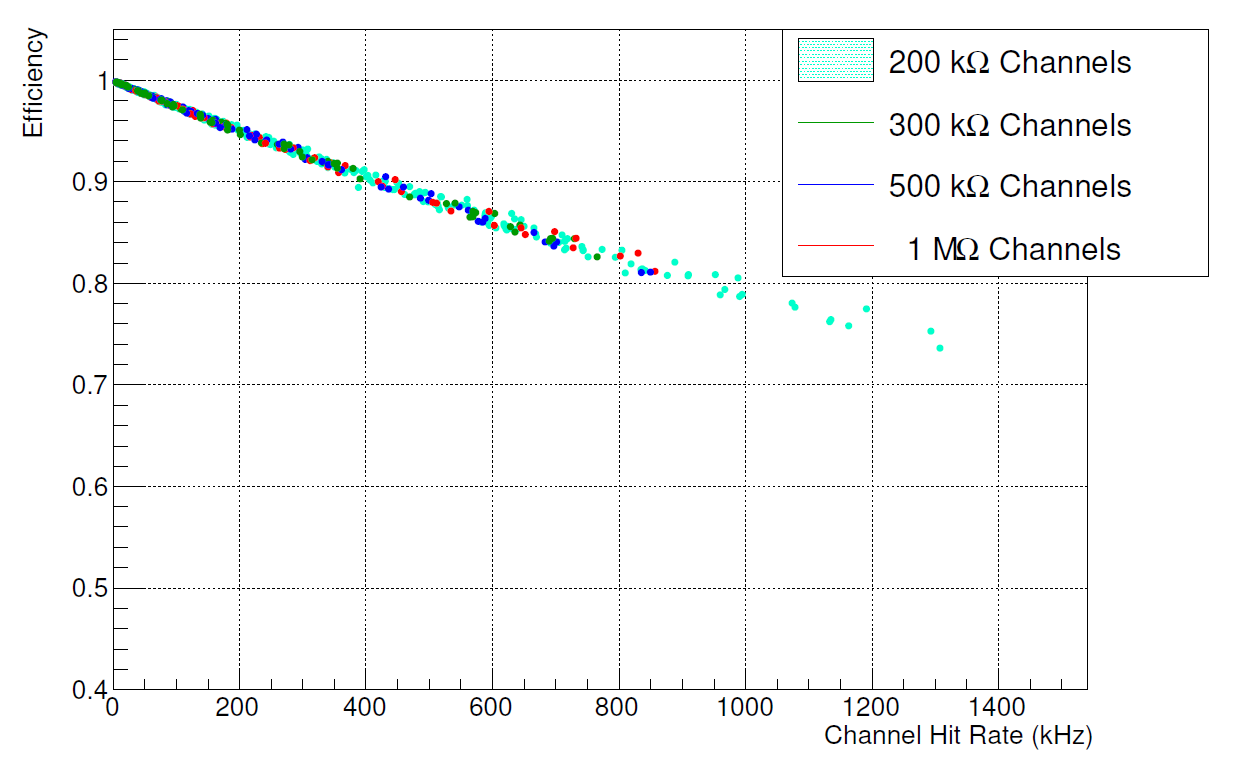}
\caption{}
\label{subfig: QS2_strip_eff}
\end{subfigure}
\qquad
\begin{subfigure}[b]{0.45\textwidth}
\includegraphics[width=\textwidth]{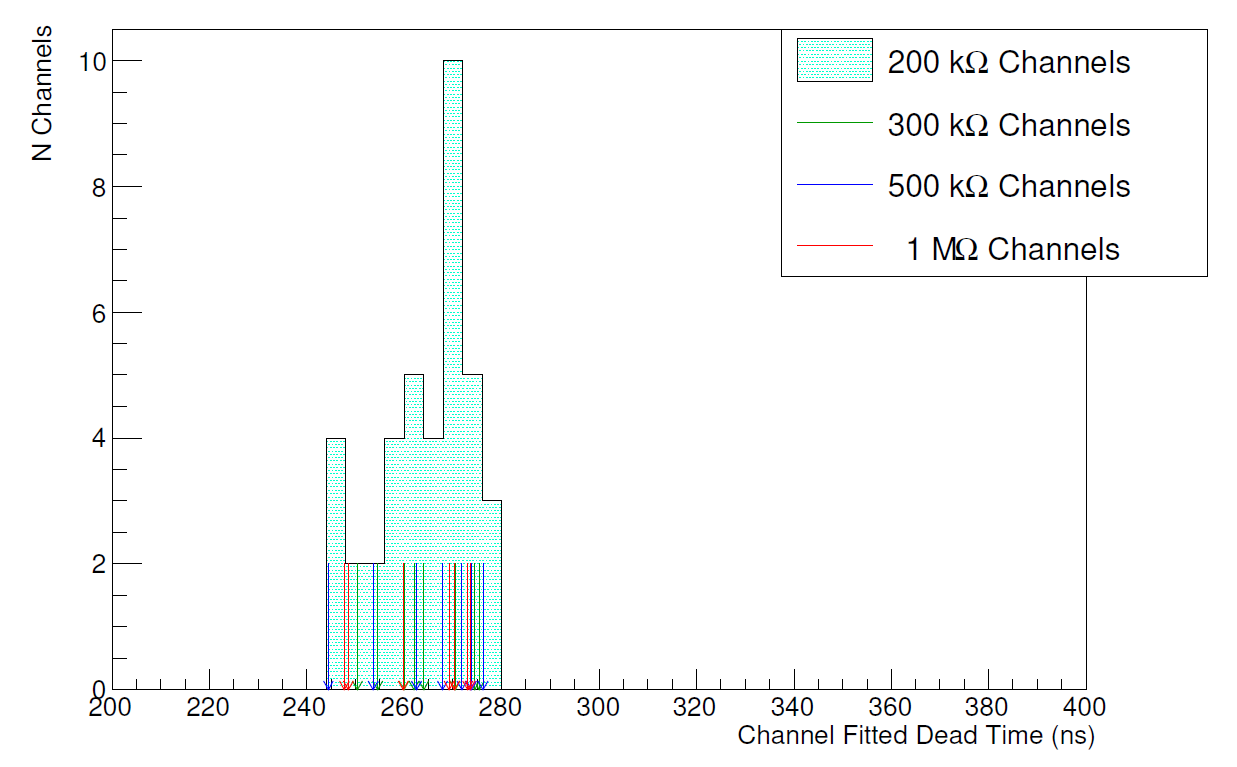}
\caption{}
\label{subfig: QS2_strip_deadtime}
\end{subfigure}
\caption{\label{fig: QS2_strip_efficiency_TP} \ref{subfig: QS2_strip_eff}: Electronics efficiency as a function of the photon hit rate for the QS2 strips with different pull-up resistor values. \ref{subfig: QS2_strip_deadtime}: Fitted dead-time for different pull-up values for the QS2 strips.}
\end{figure}

As expected, the dead-time does not change with the strip pull-up resistor value as the pull-up resistors do not attenuate the charge deposited by muon hits or background hits.  A resistor value of 400\,k$\Omega$ was chosen for the final front-end board.  The value was chosen as it decreased the compensating current from the resistor while maintaining the same dead-time.   

A shift in the baseline can be observed in strips when the pull-up resistor is 200\,k$\Omega$ and the VMM gain is increased to 3\,mV/fC from the default value of 1\,mV/fC.  The higher the gain, the more the VMM baseline recovery circuit must work in order to maintain the baseline after amplification.  In contrast, 300\,k$\Omega$, 500\,k$\Omega$ and 1\,M$\Omega$ resistors all saw no change in the baseline when the gain was increased from 1\,mV/fC to 3\,mV/fC.  Therefore the baseline recovery circuit was able to maintain the baseline even with the increased gain for these resistor values.  We chose 400\,k$\Omega$, which is greater than 300\,k$\Omega$, the smallest resistor value where no baseline shift occurred, to ensure we have some margin.


\section{Conclusion}
\label{sec: conclusion}

We performed the first direct measurement of sTGC detectors' efficiency at high background rates using the GIF++ test beam facility.  The results demonstrated that efficiency at high rates can be modeled as an exponential decay where the rate of decay is the electronics dead-time multiplied by the background rate.  The same was verified using a pulse injector emulation system which mimicked the charge deposited by the background photons in the GIF++ facility~\cite{DetEmulation}.

The measurement from both systems gave us an understanding of how the electronics dead-time changed with different parameters such as the ADC conversion time and different input charge attenuation circuits.  The operation of the full 10-bit ADC incurs a much longer dead time per hit of around 350\,ns. The faster digital mode and 6-bit ADC mode used for triggering incur short dead-times of around $220-280$ ns.

A pi-network filter is placed in front of the input on all pad channels.  The pi-network acts as a charge divider between the detector pad and the amplifier's input.  A larger attenuation factor, and therefore a smaller amount of input charge per hit, was found to also decrease the dead-time incurred. However the dead-time difference between attenuation factors of $10:1$ and $20:3$ and $5:1$ was found to be negligible in test beam data.  Emulation studies using pulse injectors confirm no real difference in dead-time for attenuation factors greater than $3:1$~\cite{DetEmulation}. The attenuation also attenuates the muon signal.  We therefore set the pi-network capacitor corresponding to the lowest attenuation factor resulting in a short electronics dead-time.  We set 200 pF, 330 pF and 470 pF as the recommended pi-network capacitance values for the final pad front-end boards. These values correspons to roughly $5:1$ attenuation factors for the Q1, Q2 and Q3 sTGC detector pads, respectively.

A pull-up resistor is placed on the input line of all sTGC strip channels.  The pull-up resistor is designed to provide a constant current that counter-acts the build up of current at the input of the ASIC at high rates due to the long signal ion tail.  No change in dead-time was observed because the pull-up resistor does not change the amount of fast electron input current which forms the prompt signal from a hit.  A 400 k$\Omega$ value was chose for the pull up resistor as it still provides enough current to counteract the long ion tail but also allows us to operate the amplifier at a gain of 3\,mV/fC.

These results not only allowed us to set the final component values used for the front-end board, but also gave us great insights into the detector and electronics performance at high rates. A better performance of the sTGC detector will make an impact in the ATLAS Level 1 Muon trigger during HL-LHC era.

\appendix
\part*{Appendices}
\addcontentsline{toc}{part}{Appendices}

\section{Expected functional form of efficiency vs hit rate}
\label{app: efficiencyVsRateTheory}

We derived in section \ref{subsec: efficiency_theory} that the relationship between efficiency and hit rate should be that of an exponential decay with the electronics dead-time $D$ times the hit rate $\lambda$ as the rate of decay.  This is defined according to equation \ref{eqn: efficiency_exponential}.  

This calculation makes the assumption that a second hit will incur additional dead-time even if it occurs within the dead-time of the first hit.  This can be seen in its asymptotic behavior at infinite rate where the measured rate, or $\epsilon \times \lambda$, would approach zero. 

Alternatively the assumption could be made that hits which arrive during dead-time will not generate dead-time of their own.  In this case, we have a dead-time of \textit{D} for every efficient hit. In that $D$ dead-time, we expect additional $D \times \lambda$ hits which do not generate dead-time. Therefore, for every one hit that is efficient, we expect one efficient hit plus $D \times \lambda$ inefficient hits in total.  Hence the efficiency is the ratio between the number of efficient hits and the number of total hits given by eq.~\ref{eqn: 1/1+x}:
 \begin{equation}
    \epsilon = \frac{1}{1+D\lambda}.
    \label{eqn: 1/1+x}
 \end{equation}
This equation now has a very different asymptotic behavior than the exponential.  The measured rate of hits, or $\lambda \times \epsilon$, approaches $1/D$ at infinite $\lambda$.

The $1/(1+D\lambda)$ calculation would be true for example if only the digital logic generated dead-time.  Hits within dead-time are never digitized, these inefficient hits would never generate dead-time in the ADC or need to wait for the digital channel to reset.  However, hits do not generate dead-time only in the digital logic, the analog amplifier and shaper must also recover back to the baseline after receiving the charge by a hit.  The higher the charge input, the longer it takes for the amplifier and shaper to recover.  It is theoretically possible that a large enough constant current into the amplifier can completely overwhelm the baseline recover circuit which can only deliver a finite amount of counter-acting current. This has not been observed in this measurements at rates that are expected for the HL-LHC.

It's important to note that both the exponential and the $1/(1+D\lambda)$ functions fits the data well for the hit rates we expect for the HL-LHC. For example, approximately 83\% efficiency at 1 MHz per channel hit rate is predicted the same both for the exponential and for the $1/(1+D\lambda$) formulas. They only noticeably differ at high background rates of 2 MHz per channel or more which is greater than the expected rate for any part of the detector. 

This is because both functional form effectively interpolate the data points.  The differences mainly occur when we extrapolate past the measured rates to greater than 3 MHz as the asymptotic behavior of the two functions are different.  Such high rates are well beyond what the sTGC expect to experience during HL-LHC runs inside ATLAS.

We choose to use the exponential functional form for this measurement.

\section{Cross-talk investigations}
\label{sec: investigations}

This appendix summarizes different studies on cross-talk effects in a sTGC detector. The focus is on effects related to the signal induction in the chamber, and not to classic "electronics cross-talk" (e.g. capacitive coupling between signal lines).
A correct estimation of the cross-talk and understanding of its causes is crucial for controlling the performance of the chambers in the ATLAS trigger system. In particular, under high intensity background the degradation of the capability to correctly reconstruct muon tracks can be severely affected. 

The sTGC QL1 quadruplet detector was used.
Muon beam and cosmic rays tests were performed, using scintillators in coincidence to select a small trigger area. The detector was also tested in the GIF++ facility, providing intense photon background.
The detector analog signal recorded by the VMM could be monitored on the oscilloscope, one channel per chip, i.e. one channel per sTGC layer. The 10-bit digitized signals from all the channels could be read out via the miniDAQ system.
The entire system is described in detail in section~\ref{sec: intro}.  
The essence of the presented results do not depend on the particular sTGC detector operation voltage and chip settings. 

\begin{figure}[h!]
\centering
\subfloat[pad]{
\includegraphics[width=0.7\textwidth]{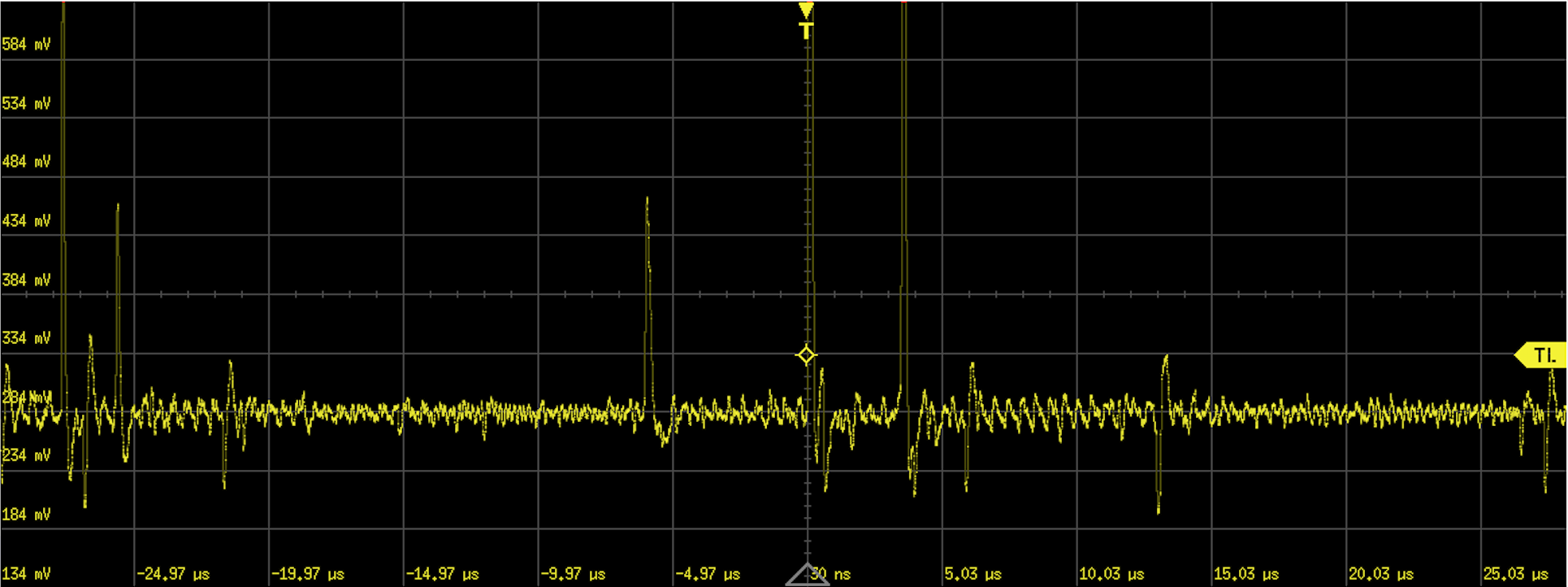}
\label{subfig: pad}
}

\subfloat[strip]{
\includegraphics[width=0.7\textwidth]{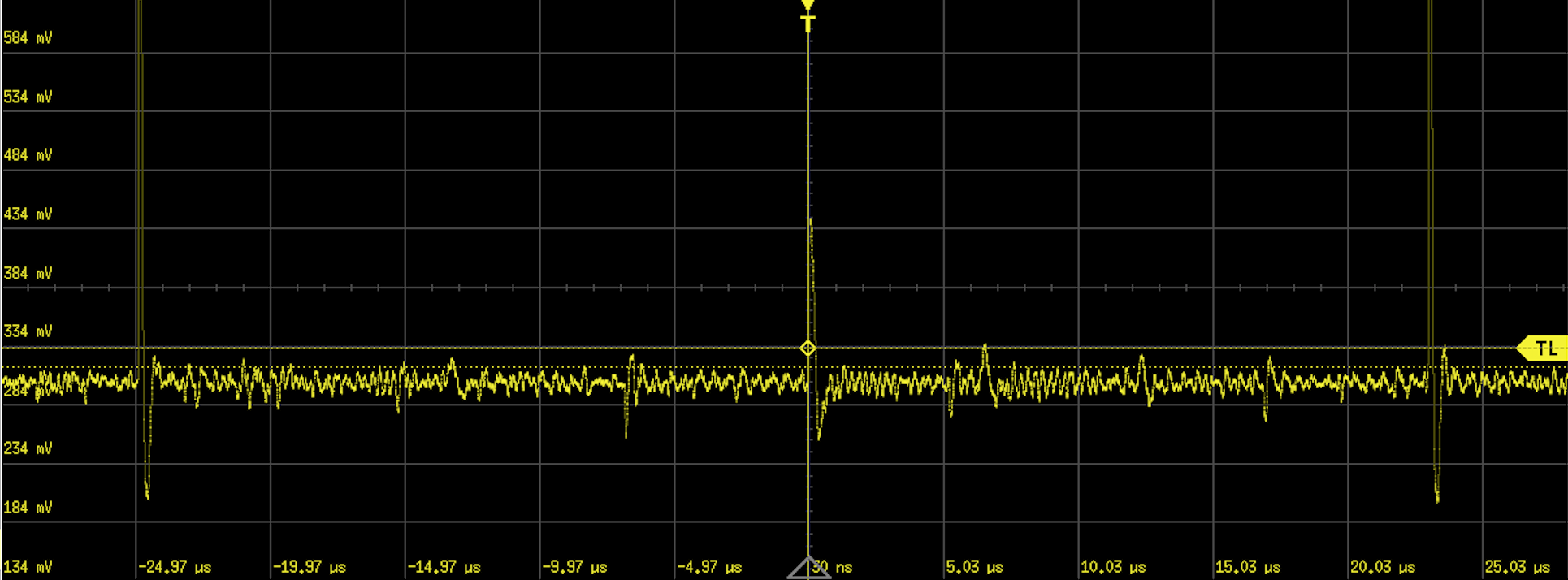}
\label{subfig: strip}
}

\subfloat[wire]{
\includegraphics[width=0.7\textwidth]{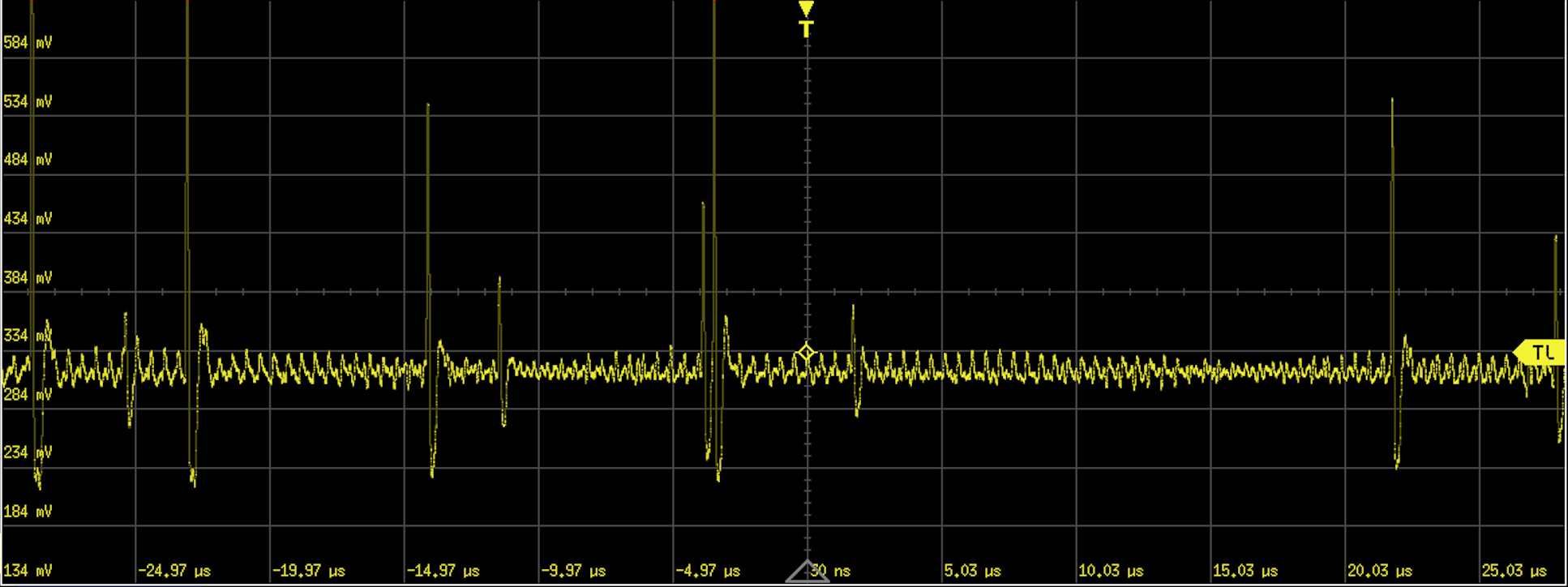}
\label{subfig: wire}
}
\caption{VMM analog readout of photon-induced signals, recorded for a sTGC pad (figure~\ref{subfig: pad}), a strip (figure~\ref{subfig: strip}), and a wire (figure~\ref{subfig: wire}).}
\label{fig: GIF++ analog signals}
\end{figure}

Figure~\ref{fig: GIF++ analog signals} presents a series of random sTGC signals recorded from a specific pad (figure~\ref{subfig: pad}), a strip (figure~\ref{subfig: strip}), and a wire (figure~\ref{subfig: wire}) of one layer of the quadruplet. In this measurement the detector was entirely irradiated by a flux of photons in the GIF++ facility. It can be noticed that in the case of pad and strip there are two distinct signal shapes. The expected one, which is a positive pulse followed by a small undershoot, and a second type, which is on average smaller in amplitude, and presents a negative pulse followed by an overshoot. The latter type does not appear in the case of the wire. The overshoot of the negative pulse is the cause of an excess of photon rate in the digitized data, compared to the estimated one. The amount of extra hits decreases slightly when using a VMM shaping time of 25~ns instead of 50~ns. This can be explained by the fact that the overshoot of these extra pulses has a slower rising time than regular positive pulses, resulting in a smaller pulse height. 

\begin{figure}[h!]
\centering
\subfloat[pad]{
\includegraphics[width=0.45\textwidth]{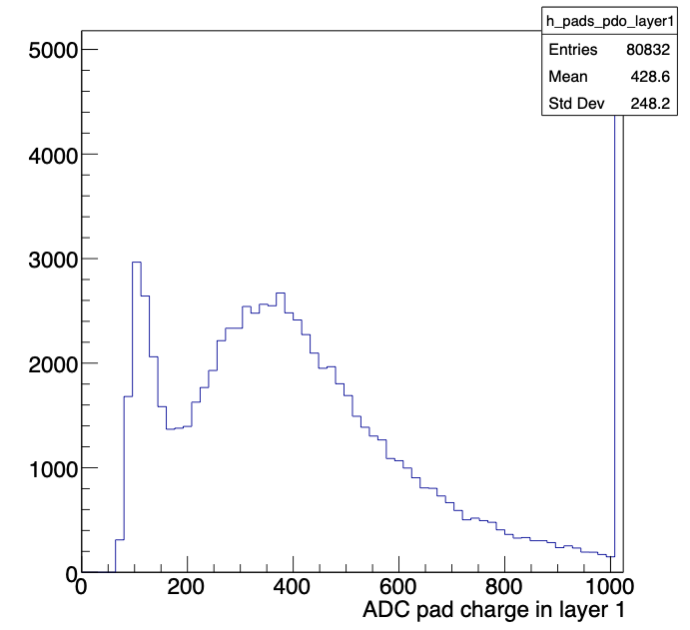}
\label{subfig: spectrum}
}
\subfloat[strip]{
\includegraphics[width=0.45\textwidth]{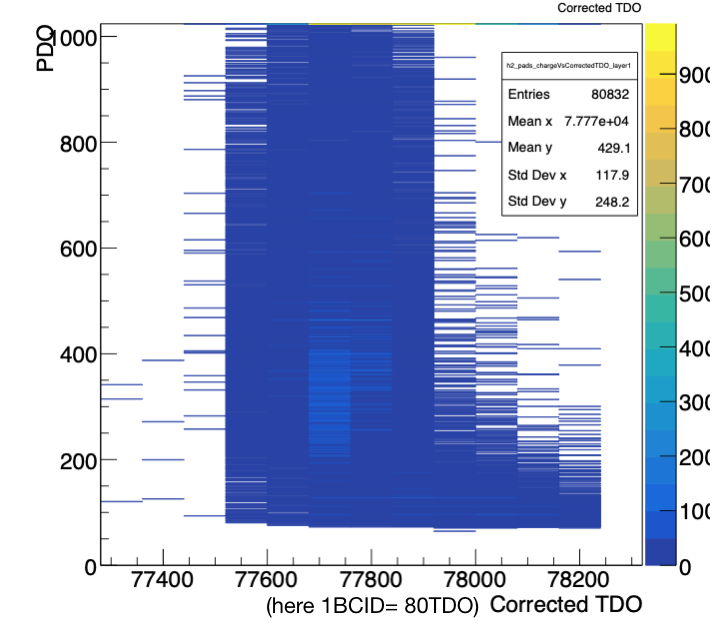}
\label{subfig: timing}
}
\caption{\ref{subfig: spectrum}: VMM 10-bit readout spectrum of muon-induced signals, recorded from sTGC pads. \ref{subfig: timing}: 10-bit charge readout as a function of the peaking time. Notice that the low charge peak is related to delayed signals.}
\label{fig: spectrum and time}
\end{figure}

\begin{figure}[h!]
\centering
\subfloat[]{
\includegraphics[width=0.45\textwidth]{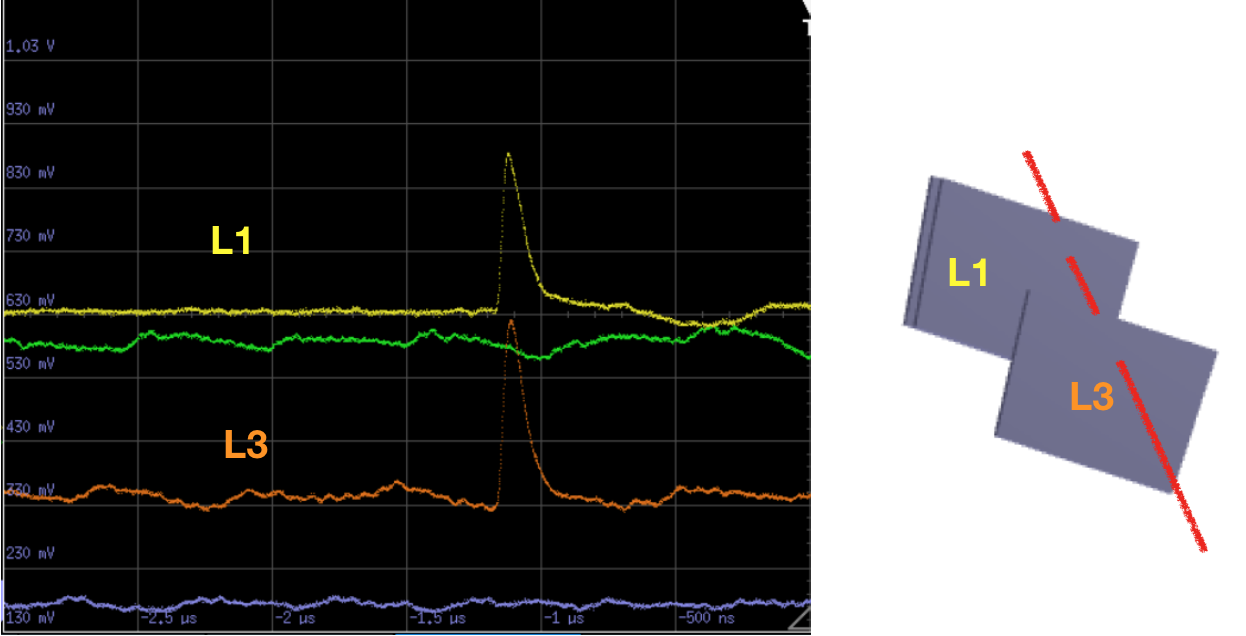}
\label{subfig: muon 1}
}
\subfloat[]{
\includegraphics[width=0.45\textwidth]{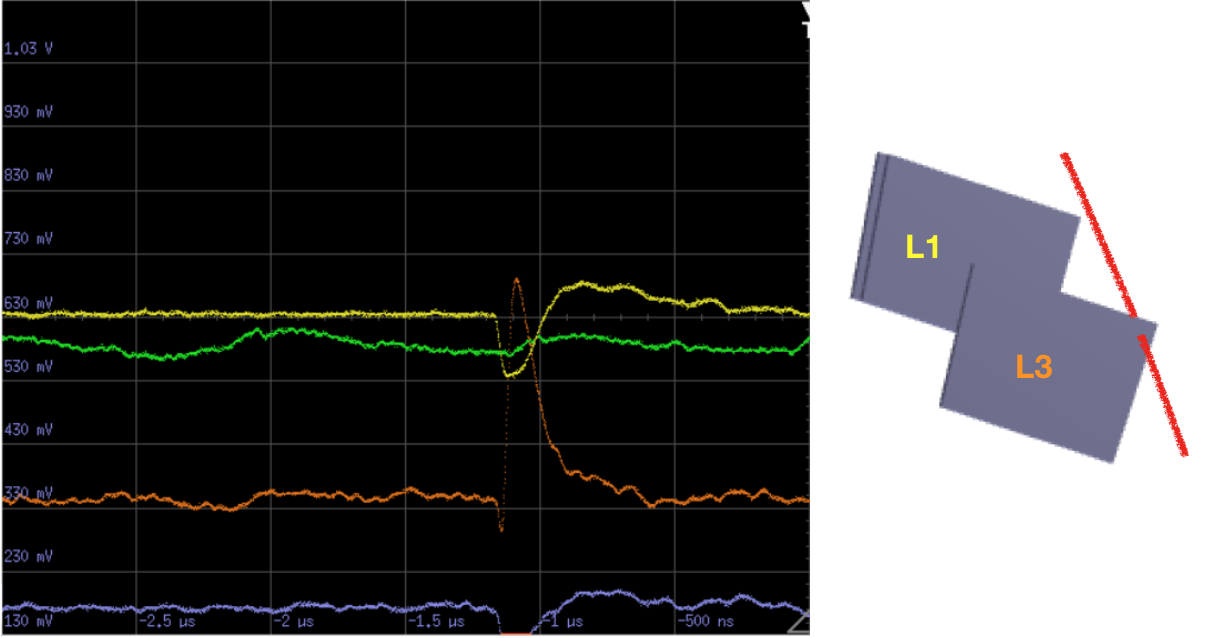}
\label{subfig: muon 2}
}

\subfloat[]{
\includegraphics[width=0.45\textwidth]{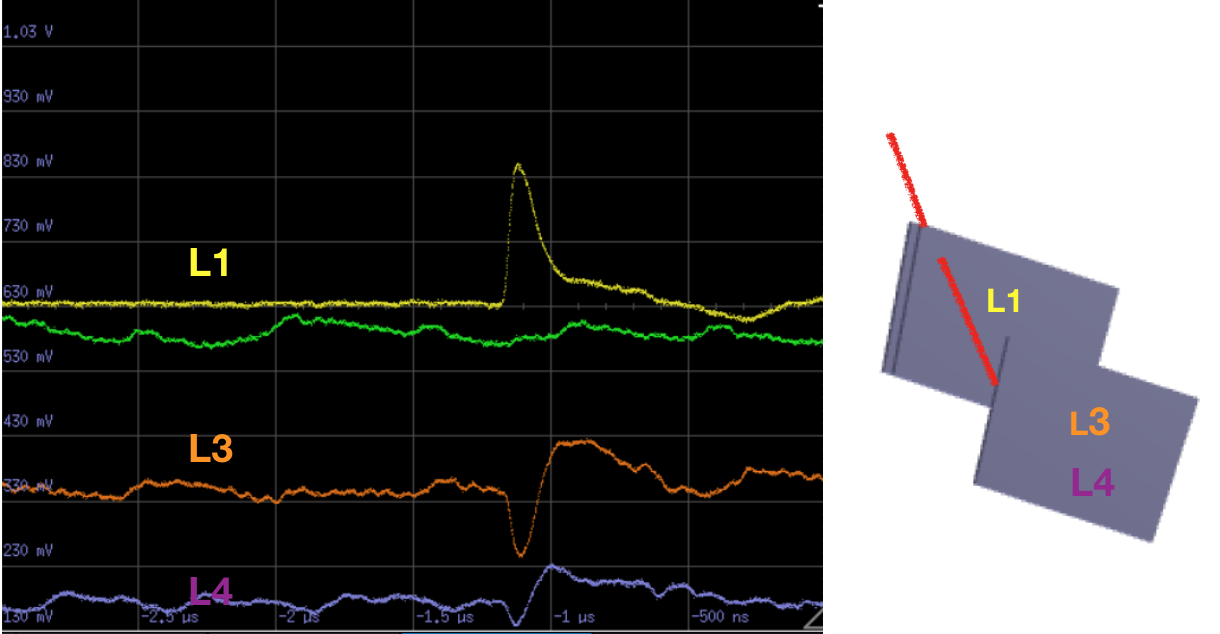}
\label{subfig: muon 3}
}
\subfloat[]{
\includegraphics[width=0.45\textwidth]{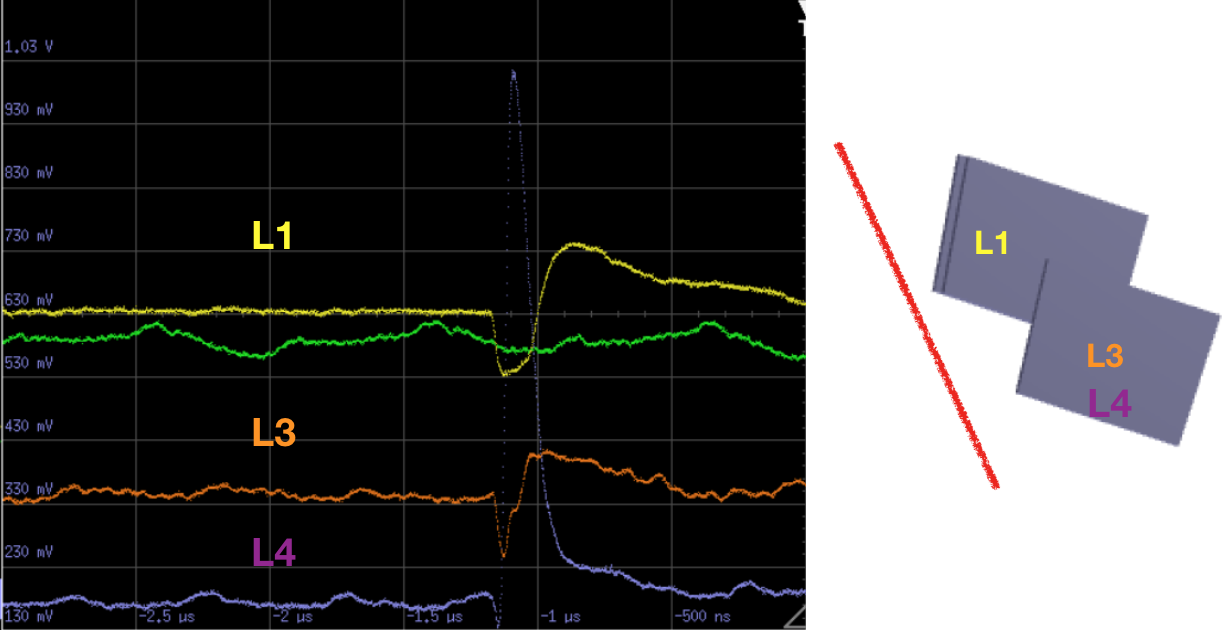}
\label{subfig: muon 4}
}

\subfloat[]{
\includegraphics[width=0.45\textwidth]{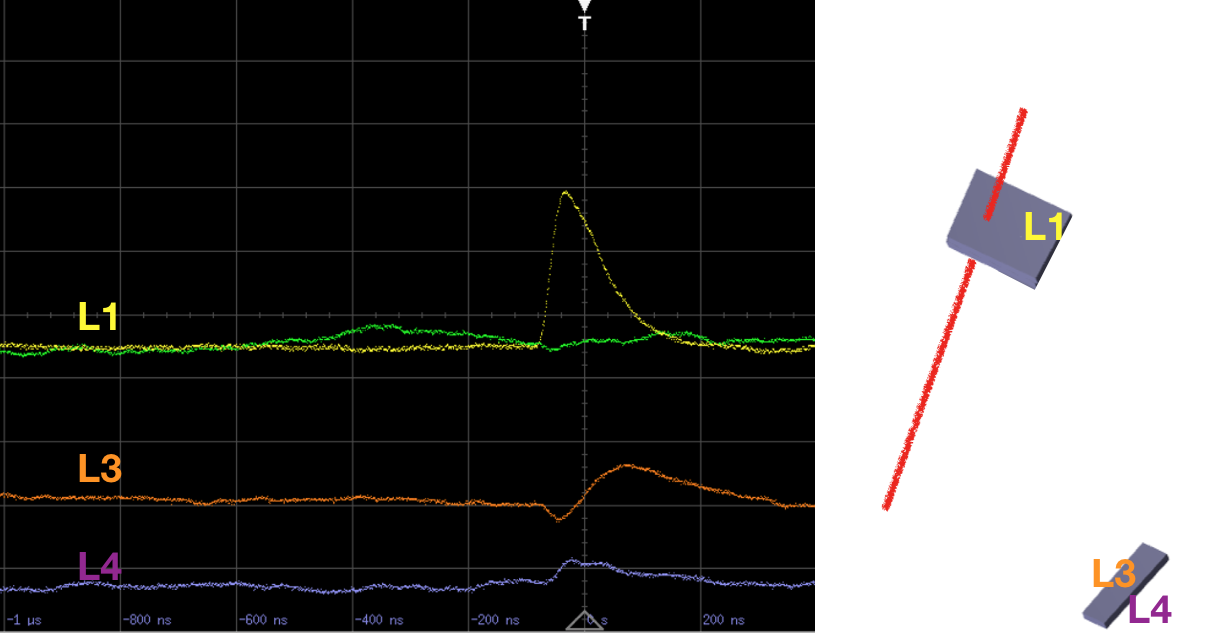}
\label{subfig: muon 5}
}
\caption{VMM analog readout of muon-induced signals, recorded from one pad for each sTGC quadruplet layer. The monitored pads are partly overlapping with the trigger region. Layer 2 was off. Layer 4 does not always appear because it was inefficient. The hypothetical track path for each event is also depicted. Notice that the overshoot of the negative peaks is delayed.}
\label{fig: muon tracks}
\end{figure}

\begin{figure}[h!]
\centering
\subfloat[]{
\includegraphics[width=0.43\textwidth]{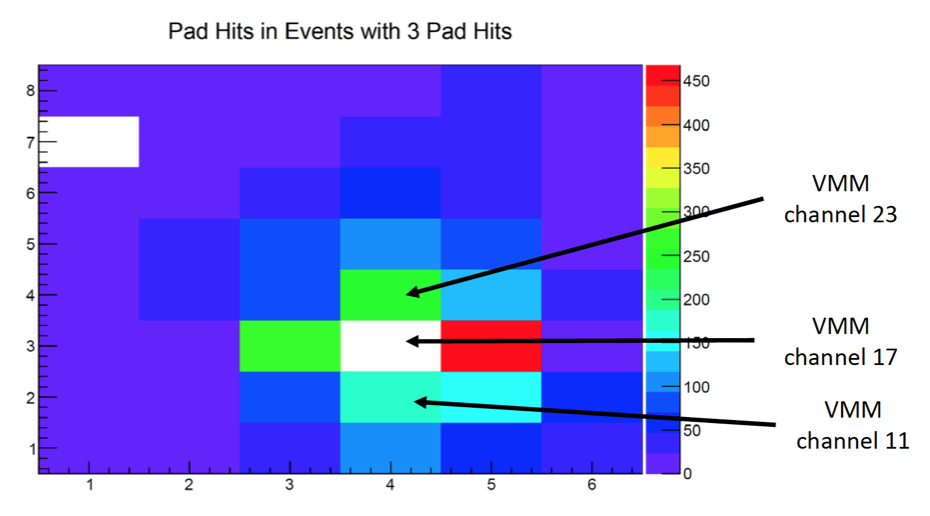}
\label{subfig: 3 hits}
}
\subfloat[]{
\includegraphics[width=0.55\textwidth]{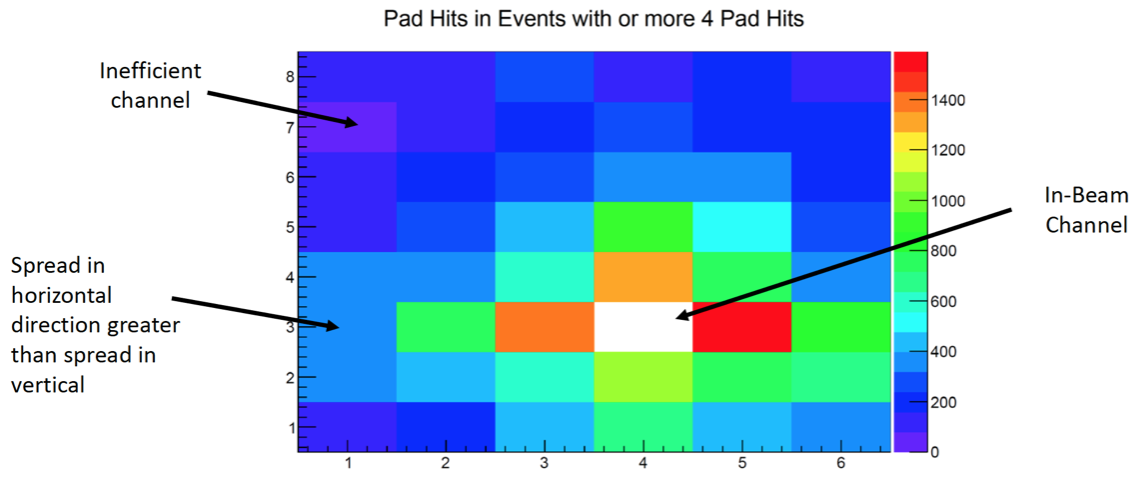}
\label{subfig: 4 hits}
}
\caption{Spacial distribution of $\upmu$ induced pad signals, recorded from a sTGC quadruplet layer. Events are selected with 3 hits~\ref{subfig: 3 hits}, and 4 hits~\ref{subfig: 4 hits}. The spread in the horizontal direction, towards the graphite ground lines, is greater than the vertical one.}
\label{fig: hits distribution}
\end{figure}

In order to understand this phenomenon in more detail, some cosmic ray tests were conducted.
In figure~\ref{fig: spectrum and time} it is shown that the charge distribution for a single pad, and also as a function of the peaking time. The threshold was set above the noise level. The lowest charge peak in figure~\ref{subfig: spectrum} is originated by the overshoot of the negative pulses, which result in a delayed peaking time, as seen from figure~\ref{subfig: timing}.
We also recorded muon-induced signals by monitoring the analog output from one pad for each sTGC quadruplet layer, in the case that the pads were partly overlapping with the trigger region and in the case that one of the pads was far away from it. 
Some representative events are shown in figure~\ref{fig: muon tracks} (mind that layer 2 was off and layer 4 was inefficient). Notice that the overshoot of the negative peaks is delayed, agreeing with figure~\ref{subfig: timing}.
Based on the fact that the negative pulses are not seen in the wires readout (see figure~\ref{subfig: wire}), and that all combinations of positive and negative pulses in coincidence are manifest, we think that the negative pulses are originated by charge dispersion on the graphite layer, reaching even very far away from the muon track (see figure~\ref{subfig: muon 5}). According to this, the hypothetical track path for each event is also depicted.
Another confirmation of the charge spread on the graphite layer as the cause of cross-talk comes from the spacial distribution of hits in events with multiple hits.
The spacial distribution of muon-induced pad signals, recorded from a sTGC quadruplet layer, are shown in figure~\ref{fig: hits distribution} for events with three hits (figure~\ref{subfig: 3 hits}), and four hits (figure~\ref{subfig: 4 hits}). The spread is in all directions, with an unbalance in the horizontal direction, towards the graphite ground lines, which are placed on both edges of the chamber. It is important to notice that there is no correspondence between the pattern and the VMM channels, suggesting that the cross-talk is happening on-detector.

\clearpage

\acknowledgments
\addcontentsline{toc}{part}{Acknowledgements}
The authors would like to thank the members of the NSW collaboration for their contributions and also the GIF++ team for supporting the gas system. We acknowledge the support of the Natural Sciences and the Engineering Research Council (NSERC) of Canada and the U.S. Department of Energy under contract DE-AC02-98CH10886.




\begin{thebibliography}{99}






\bibitem{GIF++}
D. Pfeiffer et al., \emph{The radiation field in the Gamma Irradiation Facility GIF++ at CERN}, \emph{NIM-A} 866 (2017) 91, issn: 0168-9002, url: https://www.sciencedirect.com/science/article/pii/S0168900217306113 (cit. on p. 11).

\bibitem{HitRate}
ATLAS Collaboration, \emph{Performance of the ATLAS RPC detector and Level-1 muon barrel trigger at $\sqrt(s)=13$ TeV}, \emph{JINST} {\bf 16 } (2021) P07029,doi:10.1088/1748-0221/16/07/P07029.

\bibitem{SONY_ASD}
O. Sasaki and M. Yoshida, \emph{ASD IC for the thin gap chambers in the LHC Atlas Experiment}, \emph{IEEE Transactions on Nuclear Science} 46 (1999) 1871 (cit. on p. 11).

\bibitem{DetEmulation} 
L. Moleri, N. Lupu, A. Vdovin and E. Kajomovitz, \emph{A detector-emulation method for realistic readout-electronics tests. A case study of VMM3a ASIC for sTGC detector.}, \emph{JINST} {\bf 17} (2022) P02037, url: https://doi.org/10.1088/1748-0221/17/02/p02037 (cit. on pp. 6, 7, 9, 17, 23, 25).

\bibitem{Photon_charge}
V. Smakhtin, \emph{Testing TGC detectors in a high rate environment}, \emph{NIM-A} {\bf 494} (2002) 500, Proceedings of the 8th International Conference on Instrumentation for Colliding Beam Physics, issn: 0168-9002, url: https://www.sciencedirect.com/science/article/pii/S0168900202015395 (cit. on p. 11)

\bibitem{NSW_TDR}
ATLAS Collaboration, New Small Wheel Technical Design Report, ATLAS-TDR-020.

\bibitem{sTGC_Proceeding}
A. Canesse, Small-Strip Thin Gap Chambers for the Muon Spectrometer Upgrade of the ATLAS Experiment, PoS LHCP2020 (2021) 245, ed. by B. Mansoulie, G. Marchiori, R. Salern and T. Bos (cit. on p. 4).


\bibitem{VMM_Spec}
G.  de Geronimo et al., \emph{The VMM3a ASIC}, \emph{IEEE Transactions on Nuclear Science} {\bf 69} (2022) pp. 976-985, doi: 10.1109/TNS.2022.3155818

\bibitem{ROC_Spec}
R-M. Coiban et al., \emph{The Read Out Controller for the ATLAS New Small Wheel}, \emph{JINST} {\bf 11} (2016) c02069, doi:10.1088/1748-0221/11/02/c02069

\bibitem{TDS_Spec}
J. Wang et al., \emph{Design of a Trigger Data Serializer ASIC for the Upgrade of the ATLAS Forward Muon Spectrometer}, \emph{IEEE Transactions on Nuclear Science}, {\bf 64} (2017) pp. 2958-2965, doi: 10.1109/TNS.2017.2771266.

\bibitem{SCA_Manual}
A. Caratelli et al., \emph{The {GBT}-{SCA}, a radiation tolerant {ASIC} for detector control and monitoring applications in {HEP} experiments}, \emph{JINST} {\bf 10} (2015) C03034, doi:10.1088/1748-0221/10/03/c03034

\bibitem{sTGCFEB}
P. Miao et al., \emph{The development of the Front-End Boards for the small-strip Thin Gap Chambers detector system of the {ATLAS} Muon New Small Wheel upgrade}, \emph{JINST} {\bf 15} (2020) P11024, doi:10.1088/1748-0221/15/11/p11024

\bibitem{L1DDC}
P. Gkountoumis, \emph{Level-1 Data Driver Card of the ATLAS New Small Wheel upgrade}, \emph{IEEE Nuclear Science Symposium and Medical Imaging Conference (NSS/MIC)} (2015) pp. 1-4, doi: 10.1109/NSSMIC.2015.7581785.

\bibitem{FEAST}
Faccio, Federico et al., \emph{Development of custom radiation-tolerant DCDC converter ASICs}, \emph{Journal of Instrumentation - J INSTRUM} {\bf 5} (2010). doi:10.1088/1748-0221/5/11/C11016. 

\bibitem{FELIX}
Wu, W., \emph{FELIX: The New Detector Interface for the ATLAS Detector}, \emph{IEEE Transactions on Nuclear Science} {\bf 66} (2019) pp. 986-992. doi:10.1109/TNS.2019.2913617

\bibitem{Trigger}
P. Moschovakos, "Trigger and readout electronics for the phase-I upgrade of the ATLAS forward muon spectrometer," \emph{6th International Conference on Modern Circuits and Systems Technologies (MOCAST)} (2017) pp. 1-4, doi: 10.1109/MOCAST.2017.7937658.

\bibitem{Router}
J. Wang et al., \emph{FPGA Implementation of a Fixed Latency Scheme in a Signal Packet Router for the Upgrade of ATLAS Forward Muon Trigger Electronics}, \emph{IEEE Transactions on Nuclear Science} {\bf 62} (2015) pp. 2194-2201, doi: 10.1109/TNS.2015.2477089.

\bibitem{Shockley}
W. Shockley, \emph{Currents to Conductors Induced by a Moving Point Charge},
\emph{Journal of Applied Physics} {\bf 9} (1938) pp. 635-636, doi: 10.1063/1.1710367

\bibitem{Ramo} 
S. Ramo, \emph{Currents Induced by Electron Motion}, \emph{Proceedings of the IRE}, {\bf 27}, pp. 584-585, (1939), doi: 10.1109/JRPROC.1939.228757.


\end{thebibliography}
\end{document}